\documentclass[journal=jacsat,manuscript=article]{achemso}
\usepackage{algpseudocode}%
\usepackage{listings}%
\usepackage{nicefrac}
\usepackage{physics}
\newcommand{\didv}{{d}\textit{I}/{d}\textit{V}}

\usepackage{mathrsfs}%
\usepackage{epstopdf}
\usepackage{balance}
\usepackage{comment}
\usepackage{textgreek}
\setkeys{acs}{doi = true}
\usepackage{hyperref} 
\hypersetup{
    colorlinks=true,
    linkcolor=black,      
    citecolor=black,      
    urlcolor=black,       
    filecolor=black,      
    menucolor=black,      
    runcolor=black,       
}
\usepackage{xr-hyper}
\usepackage{xr}

\makeatletter
\def\fnum@figure{\textbf{Figure~\thefigure}}
\makeatother

\author{Nicol\`{o} Bassi}
\altaffiliation{These authors contributed equally to this work}
\affiliation{nanotech@surfaces Laboratory, Empa---Swiss Federal Laboratories
for Materials Science and Technology, 8600 Dübendorf,
Switzerland}

\author{Shantanu Mishra}
\altaffiliation{These authors contributed equally to this work}
\affiliation
{nanotech@surfaces Laboratory, Empa---Swiss Federal Laboratories
for Materials Science and Technology, 8600 Dübendorf,
Switzerland}
\alsoaffiliation{Department of Physics and Astronomy, Chalmers University of Technology,
412 96 Gothenburg, Sweden}
\email{shantanu.mishra@chalmers.se}

\author{Zheng Zhang}
\affiliation{Organic and Carbon Nanomaterials Unit, Okinawa Institute of Science and Technology Graduate University, 904-0495 Okinawa, Japan}

\author{Xiao-Ye Wang}
\affiliation{Max Planck Institute for Polymer Research, 55128 Mainz, Germany}
\alsoaffiliation{College of Chemistry, Nankai University, 300071 Tianjin, P.R. China}

\author{Feifei Xiang}
\affiliation{nanotech@surfaces Laboratory, Empa---Swiss Federal Laboratories
for Materials Science and Technology, 8600 Dübendorf,
Switzerland}

\author{Nils Krane}
\affiliation{nanotech@surfaces Laboratory, Empa---Swiss Federal Laboratories
for Materials Science and Technology, 8600 Dübendorf,
Switzerland}

\author{Carlo A. Pignedoli}
\affiliation{nanotech@surfaces Laboratory, Empa---Swiss Federal Laboratories
for Materials Science and Technology, 8600 Dübendorf,
Switzerland}

\author{Klaus M\"ullen}
\affiliation{Max Planck Institute for Polymer Research, 55128 Mainz, Germany}

\author{Pascal Ruffieux}
\affiliation{nanotech@surfaces Laboratory, Empa---Swiss Federal Laboratories
for Materials Science and Technology, 8600 Dübendorf,
Switzerland}
\email{pascal.ruffieux@empa.ch}

\author{Akimitsu Narita}
\affiliation{Organic and Carbon Nanomaterials Unit, Okinawa Institute of Science and Technology Graduate University, 904-0495 Okinawa, Japan}
\alsoaffiliation{Max Planck Institute for Polymer Research, 55128 Mainz, Germany}
\email{akimitsu.narita@oist.jp}

\author{Roman Fasel}
\affiliation{nanotech@surfaces Laboratory, Empa---Swiss Federal Laboratories
for Materials Science and Technology, 8600 Dübendorf,
Switzerland}
\alsoaffiliation{Department of Chemistry, Biochemistry and Pharmaceutical Sciences, University of Bern, 3012 Bern, Switzerland}

\title
  {One-dimensional carbon nanostructures with periodic graphitic nitrogen substitution}

\begin{document}

\clearpage

\begin{abstract}

Heteroatom substitution is a powerful route to tune the chemical and electronic properties of carbon nanomaterials. In particular, replacement of an sp$^2$-hybridized carbon atom in the graphene lattice with a nitrogen atom (denoted as graphitic nitrogen) induces substantial changes in the electronic properties. These include changes in the band structure that can influence electronic transport, and magnetism. A key requirement for applications is both the periodic and precise incorporation of the heteroatoms in extended carbon lattices. Here, we report the on-surface synthesis and characterization of two one-dimensional carbon nanostructures—a polymer and a graphene nanoribbon—consisting of periodically incorporated graphitic nitrogen atoms. The on-surface reactions toward formation of the nanostructures were monitored by scanning tunneling microscopy. The bond-resolved chemical structures of the reaction intermediates and products were investigated by atomic force microscopy, which enabled atomic-scale visualization of the graphitic nitrogen sites. The electronic properties of the nanostructures were studied by scanning tunneling spectroscopy and density functional theory calculations. Our analyses revealed the presence of localized nitrogen-centered electronic states. In the gas phase where the nanostructures are in a neutral charge state, these states undergo spin polarization leading to an open-shell ground state. Upon adsorption on Au(111), the nanostructures exhibit electron transfer to the surface, which resulted in a closed-shell ground state. Our results demonstrate a straightforward and generally applicable route to synthesize graphitic nitrogen-substituted carbon nanomaterials with potential applications in spintronics, catalysis and energy storage.

\end{abstract}

\clearpage

\section{Introduction}

Heteroatom substitution offers a versatile approach to tailor the chemical and electronic properties of carbon nanomaterials. Nitrogen has been extensively studied as a substituent, particularly in the context of graphene. Following carbon in the periodic table, nitrogen shares characteristics such as similar atomic radius and the possibility to exhibit three-fold coordination with neighboring carbon atoms in the graphene lattice via sp$^2$ hybridization~\cite{guo2010controllable,joucken2015charge}. Owing to an additional valence electron in nitrogen compared to carbon, substitution of a three-fold coordinated carbon atom with nitrogen in graphene (the substitutional nitrogen atom is referred to as graphitic nitrogen, denoted hereafter as gN) results in a shift of the Fermi level away from the Dirac point, and leads to an n-type doping effect analogous to inorganic semiconductors~\cite{joucken2019electronic,joucken2012localized,ketabi2016tuning}. Beyond electronic transport, gN substitution in (nano)graphene has also been explored as a route to imprint functionalities suitable for spintronics, catalysis, and energy storage~\cite{wang2012review,da2021electronic,shao2010nitrogen,deng2016review,qu2010nitrogen}.

The proposed applications of gN-substituted carbon nanomaterials (particularly those pertaining to magnetism) require the periodic and deterministic placement of individual nitrogen atoms in carbon lattices. Conventional methods such as chemical vapor deposition and ion implantation inherently yield random distribution of nitrogen atoms and high defect densities~\cite{zhao2012production,friedman2016electronic,willke2015doping}. In this regard, on-surface synthesis provides a facile route toward the atomically precise placement of heteroatoms in carbon lattices~\cite{clair2019controlling,cai_atomically_2010,houtsma2021atomically}. Recent works have shown the on-surface synthesis of polycyclic conjugated hydrocarbons and graphene nanoribbons (GNRs) containing gN, though most examples feature pyrazine moieties that consist of two nitrogen atoms at diametrically opposite vertices in a six-membered heterocycle, and whose properties differ from isolated gN atoms~\cite{wang2017exploration,piskun2020covalent,sun2021heterocyclic,wen2022magnetic,wen2023fermi,fu2025building,sun2025surface}. The existing literature on on-surface synthesis of polycyclic frameworks with isolated gN atoms is limited to small molecules~\cite{wang2022aza,vilas2023surface,calupitan2023emergence,vegliante2025surface,li2024surface,biswas2021surface,zhang2024synthesis,ZhiHao}, or GNRs with random gN substitution and/or limited lengths~\cite{gao2022selective,bassi2024preferential}.

In this work, we demonstrate the on-surface synthesis of one-dimensional carbon nanostructures with periodic gN substitution on Au(111), and their characterization at the atomic scale by scanning tunneling microscopy (STM), atomic force microscopy (AFM), and density functional theory (DFT) calculations. Specifically, we targeted the synthesis of a polymer and a chevron GNR based on covalently coupled hexabenzocoronene (HBC)-like building blocks, each containing a single gN atom (Figure~\ref{OSS}). We characterized the on-surface reaction steps and the bond-resolved chemical structures of the polymer (denoted \textit{poly}-gN-HBC) and the GNR (denoted GNR-gN-HBC) by STM and AFM measurements at a temperature of 5~K. We probed the electronic properties of \textit{poly}-gN-HBC and GNR-gN-HBC by scanning tunneling spectroscopy (STS) measurements. We detected and mapped the distribution of frontier electronic bands associated with gN substitution in real space (that indicate substantial modification of the electronic properties compared to the corresponding all-carbon counterparts), along with the bands associated with the carbon framework. In the gas phase, DFT calculations predicted an open-shell antiferromagnetic ground state of \textit{poly}-gN-HBC and GNR-gN-HBC. However, comparison of STS measurements with DFT calculations indicated that each gN-HBC unit transferred an electron to Au(111), which resulted in a closed-shell ground state of \textit{poly}-gN-HBC and GNR-gN-HBC in their cationic state.

\section{Results and discussion}

\subsection{On-surface synthesis and structural characterization}

\begin{figure}[!h]
\begin{center}
\includegraphics[width=0.8\linewidth]{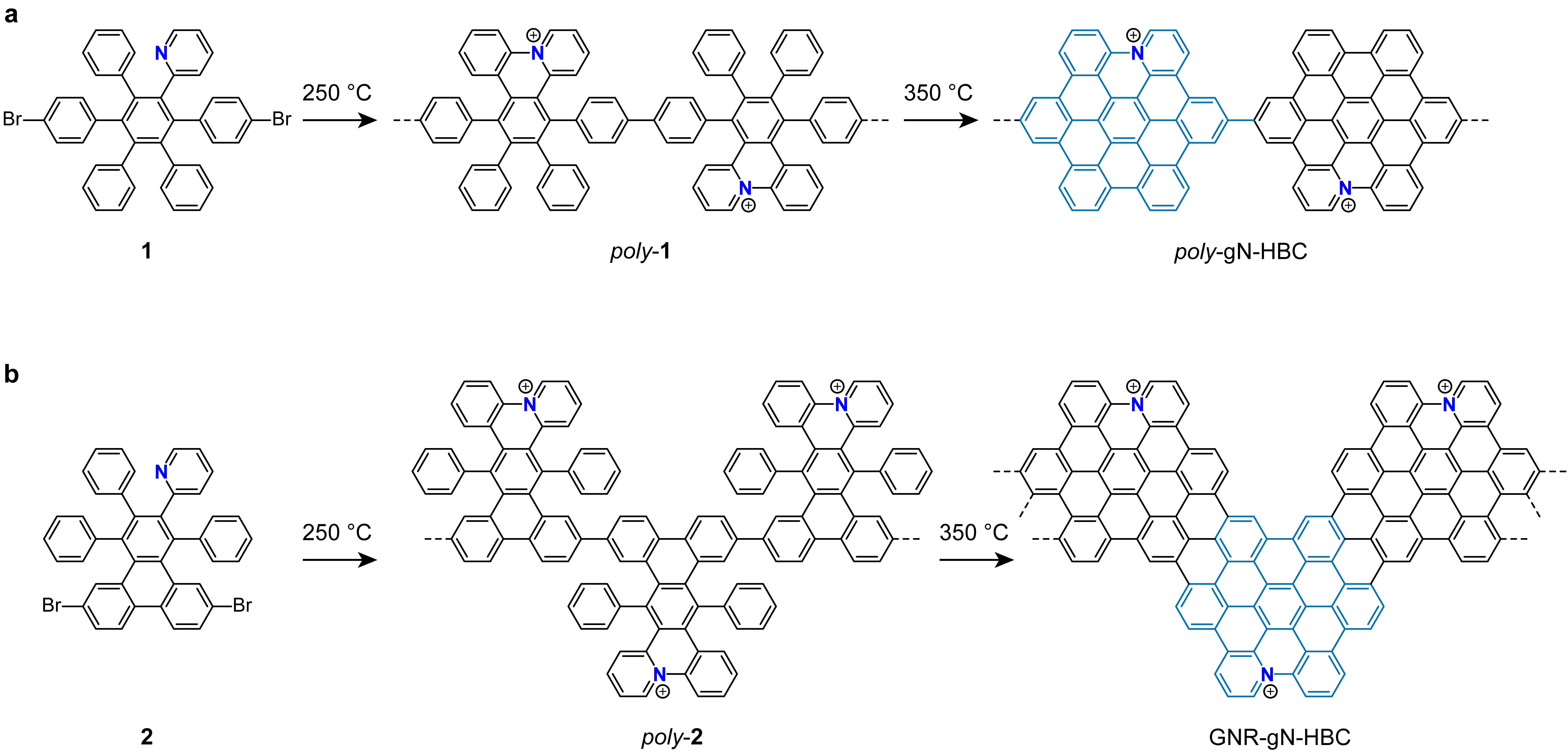}
\caption{Synthetic route toward the graphitic nitrogen-substituted nanostructures reported in this work. (a, b) Scheme of on-surface synthesis of \textit{poly}-gN-HBC (a) and GNR-gN-HBC (b) on Au(111). The gN-HBC units in \textit{poly}-gN-HBC and GNR-gN-HBC are highlighted. We note that debrominative aryl-aryl coupling and C--N bond formation occur within different temperature windows, with our experiments suggesting that the former precedes the latter (see Figure S14). While we cannot specify the exact temperature windows within which the individual reactions occur, both reactions are nearly complete at 250~$^\circ$C.}
\label{OSS}
\end{center}
\end{figure}

Our strategy for synthesizing \textit{poly}-gN-HBC and GNR-gN-HBC began with the solution-phase synthesis of the respective dibrominated precursors \textbf{1} and \textbf{2} (Figure~\ref{OSS}; see Figures~\ref{s3}--\ref{nmr6} and accompanying text for details).  Thereafter, submonolayer coverages of the precursors were deposited onto single-crystalline Au(111) surfaces held at room temperature and under ultrahigh vacuum (see Figures~\ref{320RT} and~\ref{467RT} for characterizations of the precursors). A two-step annealing first induced debrominative aryl--aryl coupling at $\sim$200~$^\circ$C that led to intermediate polymer phases, followed by cyclodehydrogenation reaction at 350~$^\circ$C that resulted in the formation of \textit{poly}-gN-HBC and GNR-gN-HBC (shown schematically in Figure~\ref{OSS})~\cite{cai_atomically_2010}. Notably, we observed cyclodehydrogenative C--N bond formation to occur at temperatures around 200~$^\circ$C, much lower than temperatures required for the corresponding C--C bond formation (typically occurring at 300~$^\circ$C or above~\cite{ooeTwoStepOnSurfaceSynthesis2023}) and in line with previous reports~\cite{bassi2024preferential,piskun2020covalent}. Figures~\ref{320poly1}--\ref{467poly1} show scanning probe and theoretical characterizations of the intermediate polymer phases.

\begin{figure}[!ht]
\begin{center}
\includegraphics[width=0.8\linewidth]{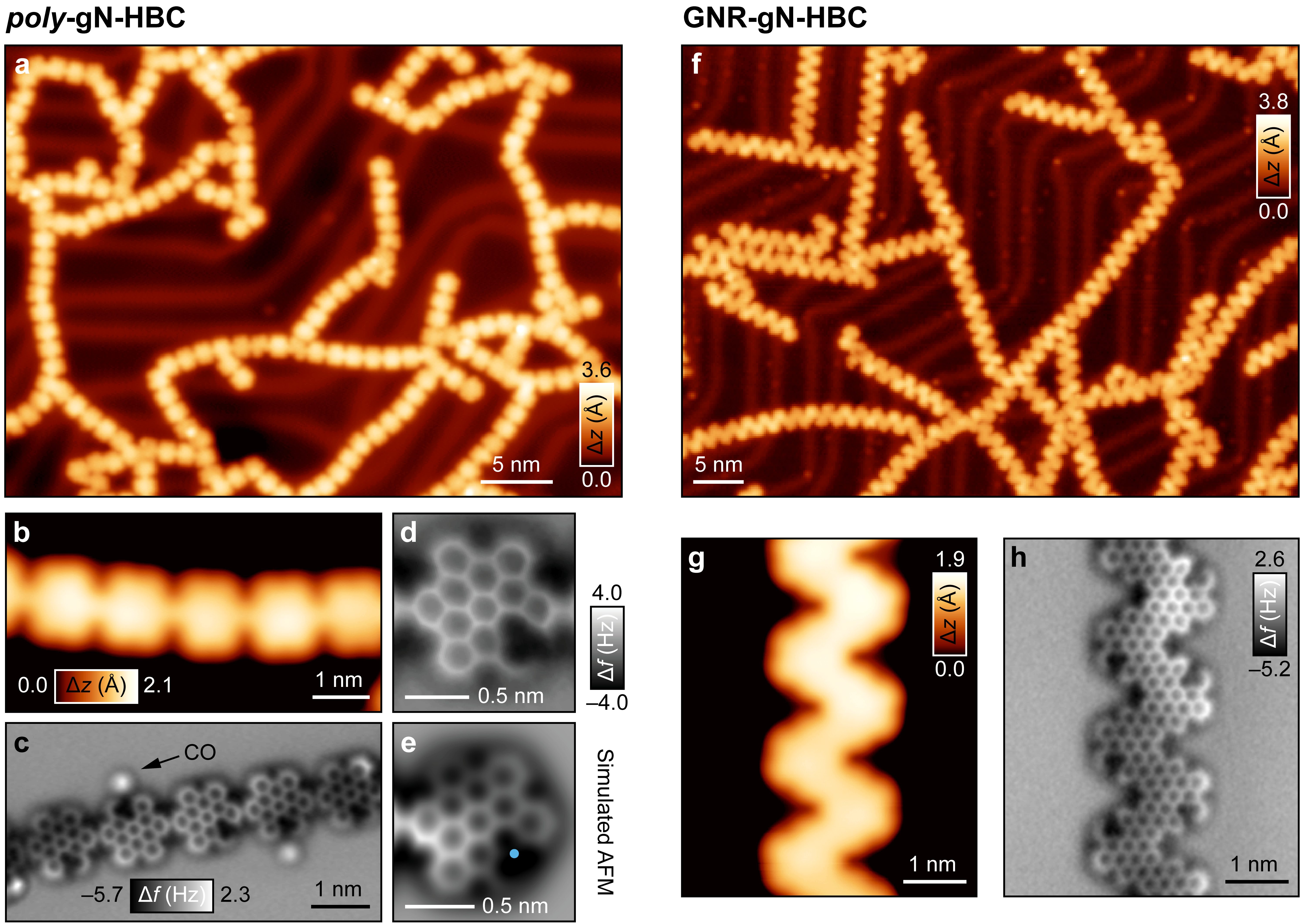}
\caption{Imaging and structural characterization of the graphitic nitrogen-substituted nanostructures. (a) STM overview image after two-step annealing of~\textbf{1} on Au(111) at 250~$^\circ$C and 350~$^\circ$C ($V = -1.0$~V, $I = 100$~pA). $\Delta z$ denotes the tip height, with positive (negative) values indicating tip retraction (approach) from the set-point conditions. (b) High-resolution STM image of a \textit{poly}-gN-HBC segment ($V = -30$~mV, $I = 120$~pA). (c) AFM image of a \textit{poly}-gN-HBC segment. $\Delta f$ denotes the frequency shift. A co-adsorbed carbon monoxide (CO) molecule is indicated by an arrow. STM set-point: $V=5$~mV and $I=100$~pA on Au(111); $\Delta z = -0.3$~\AA. (d) High-resolution AFM image of an individual gN-HBC unit in a polymer. STM set-point: $V=5$~mV and $I=100$~pA on Au(111); $\Delta z = -0.5$~\AA. The images in (c, d) were acquired with the same tip. (e) Simulated AFM image of an individual gN-HBC unit in a gN-HBC dimer. The geometry of the dimer on Au(111) was optimized using DFT. The filled circle denotes the position of the nitrogen atom.  (f) STM overview image after two-step annealing of \textbf{2} on Au(111) at 180~$^\circ$C and 350~$^\circ$C ($V = -1.5$~V, $I = 30$~pA). (g) High-resolution STM image of a GNR-gN-HBC segment ($V = -0.2$~V, $I = 100$~pA). (h) AFM image of the GNR-gN-HBC segment. STM set-point: $V=5$~mV and $I=100$~pA on Au(111); $\Delta z = -0.4$~\AA.}
\label{STMandAFM}
\end{center}
\end{figure}

Figure~\ref{STMandAFM}a shows an STM overview image of the surface after a two-step annealing of \textbf{1} on Au(111) at temperatures of 250~$^\circ$C and 350~$^\circ$C, revealing the formation of \textit{poly}-gN-HBC and highlighting the success of our synthetic strategy. High-resolution in-gap STM imaging of a \textit{poly}-gN-HBC segment (Figure~\ref{STMandAFM}b) showed an overall smooth topography of the individual gN-HBC units. To directly visualize the gN sites, we conducted AFM imaging. Figure~\ref{STMandAFM}c shows an AFM image of a \textit{poly}-gN-HBC segment, wherein the carbon framework of the gN-HBC units, the C--C bonds between the units, and a characteristic dark feature at the gN sites (in agreement with previous reports~\cite{biswas2021surface,kawai2018multiple}) are evident (see Figure~\ref{STMandAFM}d for high-resolution AFM imaging of an individual gN-HBC unit, and Figure~\ref{320a_AFM} for additional AFM data). Given the freedom of rotation of the pyridyl ring about the C--C single bond in \textbf{1}, C--N bond formation can occur either along the long axis of the polymer or at an angle of 60$^\circ$ to the long axis, which explains the varying positions of the nitrogen atoms in the segment. Simulated AFM image of a gN-HBC unit (Figure~\ref{STMandAFM}e) reproduced the features in the experimental AFM images, including the dark appearance of the gN sites.

Likewise, we extended our protocol toward the synthesis of GNR-gN-HBC. Figures~\ref{STMandAFM}f,g show overview and high-resolution STM images, respectively, after a two-step annealing of \textbf{2} on Au(111) at 180~$^\circ$C and 350~$^\circ$C, revealing the formation of nearly defect-free GNRs with maximum observed lengths of $\sim$50~nm. AFM imaging (Figure~\ref{STMandAFM}h) confirmed periodic gN incorporation in the GNRs, and similar to \textit{poly}-gN-HBC, the GNRs exhibit both on- and off-axis C--N bond formation. The successful synthesis of both \textit{poly}-gN-HBC and GNR-gN-HBC demonstrates the generality of our synthetic approach toward the formation of gN-substituted carbon nanostructures.

\begin{figure}[!ht]
\begin{center}
\includegraphics[width=0.7\linewidth]{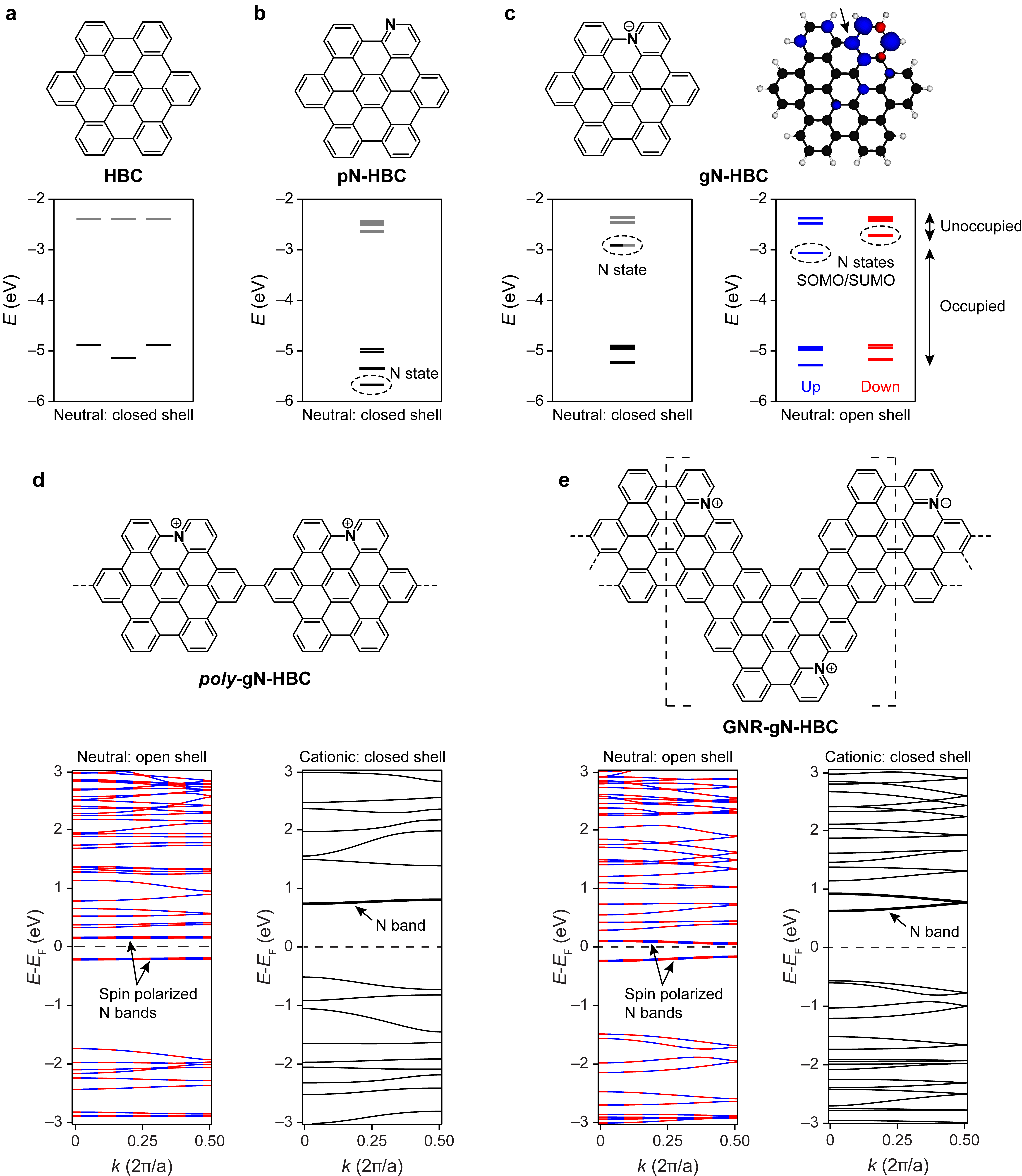}
\caption{Theoretical electronic characterization of the graphitic nitrogen-substituted nanostructures. (a, b) Chemical structure and orbital energy diagram of HBC (a) and pN-HBC (b). Occupied and unoccupied states in (a--c) are colored black and gray, respectively. (c) Chemical structure and orbital energy diagram of gN-HBC in the closed-shell state (left). Also shown are the spin-density isosurface plot of gN-HBC (isovalue: 0.003~e/Bohr$^3$; the location of the nitrogen atom is indicated by an arrow) and the orbital energy diagram in the open-shell state (right). The two colors represent spin up and spin down. SOMO (SUMO) denotes singly occupied (unoccupied) molecular orbital. The energies in (a--c) are given relative to the vacuum level. (d, e) Chemical structure and band structure plots of \textit{poly}-gN-HBC (d) and GNR-gN-HBC (e) in the neutral antiferromagnetic and cationic closed-shell states. In the open-shell state, spin-up and spin-down bands are degenerate. The energies in (d, e) are given relative to the respective Fermi energies (\textit{E}\textsubscript{F}). The unit cells for the band structure calculations correspond to the gN-HBC monomer and dimer for the cationic and neutral states of \textit{poly}-gN-HBC, respectively (d), and the area enclosed by the dashed brackets for GNR-gN-HBC (e).}
\label{elec}
\end{center}
\end{figure}

\subsection{Electronic characterization}

We first assessed the impact of nitrogen substitution on the electronic structure of all-carbon frameworks by DFT calculations, and we started by considering HBC as a model system. As shown in Figure~\ref{elec}a, the frontier electronic structure of HBC consists of a pair of degenerate orbitals constituting the highest occupied molecular orbital (HOMO) and HOMO--1, and a pair of degenerate orbitals constituting the lowest unoccupied molecular orbital (LUMO) and LUMO+1 (themselves nearly degenerate with the LUMO+2), along with a wide HOMO--LUMO gap of $\sim$2.50~eV. Replacing a =CH-- group at the periphery by a nitrogen atom (referred to as a pyridinic nitrogen, denoted pN) lifts the degeneracy of the frontier states and leads to a downward shift in the energies of the HOMO (LUMO) by 0.10~eV (0.22~eV) because of the higher electronegativity of nitrogen compared to carbon (Figure~\ref{elec}b)~\cite{bronner_aligning_2013,cai2014graphene}. Related to nitrogen substitution in pN-HBC, there is a deep-lying nitrogen state (Figure~\ref{elec}b, indicated by a dashed ellipse). In contrast, gN substitution leads to a more drastic impact on the electronic structure. Whereas pN is isoelectronic to the =CH-- group it replaces in a conjugated system, gN brings one more valence electron than the carbon atom it replaces. This additional valence electron leads to a non-bonding orbital in the gN-HBC spectrum (Figure~\ref{elec}c, left; indicated by a dashed ellipse). Inclusion of electronic correlations via spin-polarized DFT calculations leads to spin splitting of this state into singly occupied and corresponding unoccupied molecular orbitals (Figure~\ref{elec}c, right; also shown is the spin-density isosurface plot of the gN-HBC unit).

By extension, in \textit{poly}-gN-HBC, where the individual gN-HBC units are connected via C--C $\sigma$ bonds, the weak electronic coupling between neighboring units gives rise to nearly dispersionless spin-polarized frontier bands associated with nitrogen substitution, and an antiferromagnetic coupling between neighboring units (Figure~\ref{elec}d, left). Incorporating gN atoms into the rigid $\pi$ framework of GNR-gN-HBC leads to an increase in the dispersion of the spin-polarized frontier bands (Figure~\ref{elec}e, left) while keeping the ground state antiferromagnetic, which may make this system more suitable from a device point of view, particularly when spin and charge transport is desired. We note that the results of our calculations agree with a previous theoretical work by Azev\^{e}do et al.~\cite{da2021electronic}. Relative to the antiferromagnetic ground state, the energy of the ferromagnetic (closed-shell) state is 6 meV (130 meV) and 32 meV (62 meV) higher for \textit{poly}-gN-HBC and GNR-gN-HBC, respectively (see Figure~\ref{spin polari} for additional DFT calculations).

At the outset, we note that as opposed to their neutral charge states in the gas phase, both \textit{poly}-gN-HBC and GNR-gN-HBC are positively charged (that is, in a cationic state) on Au(111) because of each constituent gN-HBC unit losing one electron to the surface. In such a scenario, \textit{poly}-gN-HBC and GNR-gN-HBC are no longer open-shell systems as in the neutral charge state but are closed shell. As shown in the DFT band structure calculations, the former singly occupied bands of both systems in the neutral state are emptied in the cationic state (Figures~\ref{elec}d,e, right). Consequently, the frontier bands in the cationic state correspond to a valence band (VB) with dominant contribution from the carbon atoms, and a conduction band (CB) associated with nitrogen substitution.

\begin{figure}[!h]
\begin{center}
\includegraphics[width=0.9\linewidth]{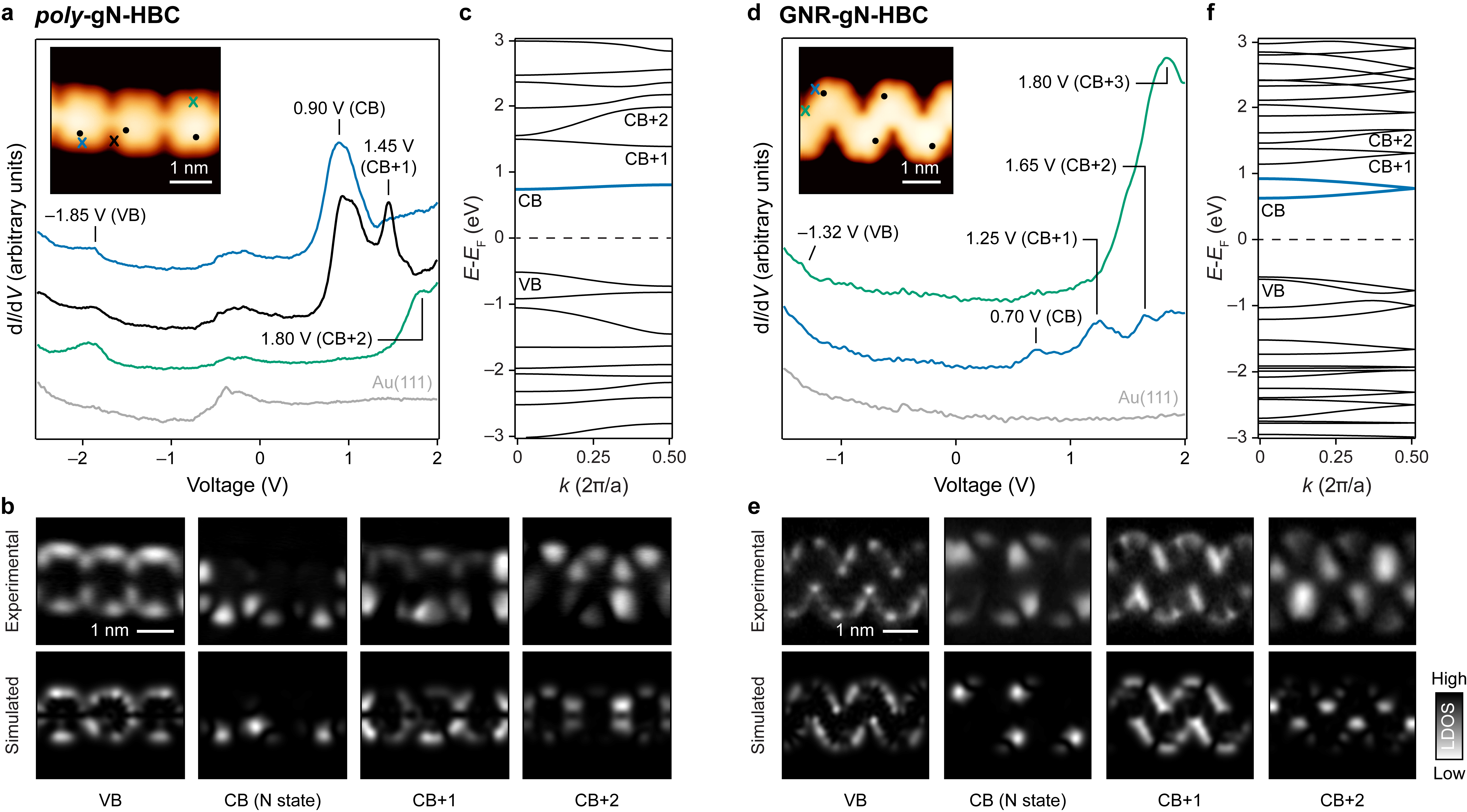}
\caption{Experimental electronic characterization of the graphitic nitrogen-substituted nanostructures. (a) \didv spectra acquired on a \textit{poly}-gN-HBC segment. Open feedback parameters: $V = -2.50$~V and $I = 200$~pA on the individual units; root-mean-square modulation voltage in the lock-in amplifier $V_{\mathrm{rms}} = 16$~mV. Acquisition positions of the spectra are indicated by the respective colored crosses in the inset STM image ($V = -30$~mV, $I = 120$~pA). Filled circles in the STM images in (a, d) denote the positions of the nitrogen atoms in each unit. (b) Constant-current \didv maps of the segment acquired at voltages corresponding to the resonances in (a) (top), and the corresponding simulated LDOS maps (bottom). Scanning parameters (from left to right): $V = -1.85$~V, $I = 250$~pA; $V = 0.70$~V, $I = 200$~pA; $V = 1.44$~V, $I = 200$~pA; and $V = 1.82$~V, $I = 200$~pA; $V_{\mathrm{rms}} = 10$~mV. (c) Band structure of cationic \textit{poly}-gN-HBC, with the bands relevant to the STS measurements labeled. (d) \didv spectra acquired on a GNR-gN-HBC segment. Open feedback parameters: $V = -2.00$~V and $I = 200$~pA ($I = 300$~pA for the blue curve); $V_{\mathrm{rms}} = 20$~mV. Acquisition positions are indicated by the respective colored crosses in the inset STM image ($V = -0.20$~V, $I = 100$~pA). (e) Constant-current \didv maps of the segment acquired at voltages corresponding to the resonances in (d) (top), and the corresponding simulated LDOS maps (bottom). Scanning parameters (from left to right): $V = -1.35$~V, $I = 200$~pA; $V = 0.70$~V, $I = 200$~pA; $V = 1.20$~V, $I = 200$~pA; and $V = 1.60$~V, $I = 200$~pA; $V_{\mathrm{rms}} = 10$~mV. (f) Band structure of cationic GNR-gN-HBC, with the bands relevant to the STS measurements labeled. The unoccupied nitrogen bands (CB) are highlighted in color and with a thicker line stroke in (c, f).}
\label{didv}
\end{center}
\end{figure}

Experimentally, we probed the electronic properties of \textit{poly}-gN-HBC and GNR-gN-HBC by STS. Figure~\ref{didv}a shows \didv spectra ($I$ and $V$ denote the tunneling current and bias voltage, respectively) acquired at various locations on a \textit{poly}-gN-HBC segment. The spectra exhibited four resonances at $V = -1.85$~V, 0.90~V, 1.45~V, and 1.80~V. To visualize the spatial distributions of the resonances, we acquired \didv maps at approximately these voltages (Figure~\ref{didv}b, top row) and compared them to the calculated local density of states (LDOS) maps of cationic \textit{poly}-gN-HBC (Figure~\ref{didv}b, bottom row; corresponding bands are labeled in Figure~\ref{didv}c). The map at $V = -1.85$~V revealed a near-uniform distribution of intensities along the periphery of the gN-HBC units (the intensities are slightly modified by the presence of gN sites) and agreed well with the calculated LDOS map at the VB maximum with a dominant contribution from the carbon sites. The state at $V = 0.90$~V manifests as prominent lobes localized at the gN sites of the individual gN-HBC units and exhibited good agreement with the calculated LDOS map at the CB minimum, corresponding to the unoccupied nitrogen band. The states at $V = 1.45$~V and 1.80~V, which are also distributed over the periphery of the gN-HBC units, are assigned to the CB+1 and CB+2, with dominant contributions from the carbon sites (see Figure~\ref{maps} for additional \didv maps).

\didv spectra acquired on a GNR-gN-HBC segment (Figure~\ref{didv}d) exhibited four frontier resonances at $V = -1.32$~V, 0.70~V, 1.25~V, and 1.65~V. The corresponding \didv maps (Figure~\ref{didv}e, top row) revealed that the state at $V = -1.32$~V is distributed over the periphery of the GNR with enhanced intensity at the gN sites, which agreed well with the calculated LDOS map (Figure~\ref{didv}e, bottom row; see Figure~\ref{didv}f for the corresponding bands) at the VB maximum, which has a dominant contribution from the carbon sites. The state at $V = 0.70$~V is localized at the gN sites and agreed well with the calculated LDOS map at the CB minimum, corresponding to the unoccupied nitrogen band. Likewise, the states at $V = 1.25$~V, distributed over the GNR periphery with particularly strong intensities along the non-nitrogen-containing oblique edges, and at $V = 1.65$~V, with localized intensities at the bay regions, correspond to the CB+1 and CB+2 of the GNR, respectively, with dominant contributions from the carbon sites. The enhanced delocalization due to the rigid $\pi$ framework of the GNR manifested in the nitrogen state in GNR-gN-HBC ($V = 0.70$~V) being closer to the Fermi energy than in \textit{poly}-gN-HBC ($V = 0.90$~V).

\section{Conclusions}

In summary, we demonstrated the on-surface synthesis of two graphitic nitrogen-substituted one-dimensional carbon nanostructures, a polymer and a graphene nanoribbon, with atomic precision and periodic nitrogen substitution. The success of our synthetic protocol, along with the high fidelity of nitrogen substitution in the nanostructures, is demonstrated via atomic-scale imaging using scanning probe techniques. Spectroscopic measurements, in combination with density functional theory calculations, revealed localized nitrogen-centered bands. In the neutral charge state of the nanostructures, the nitrogen-centered bands are spin polarized, with an antiferromagnetic coupling between neighboring nitrogen sites. We also demonstrated that the dispersion of the nitrogen-centered bands, which can influence transport, can be controlled by the design of the nanostructure framework. Our synthetic protocol can be extended for graphitic nitrogen incorporation in a range of carbon nanostructures with different structural designs and dimensionalities. This opens up a route to imprint magnetism and energy storage functionalities in organic materials.

\clearpage

\section{Methods}

\subsection{Sample preparation and scanning probe characterization}

Au(111) single crystal surfaces (MaTeck GmbH) were prepared by iterative cycles of Ar\textsuperscript{+} sputtering and annealing at 723~K. Prior to sublimation of molecules, the surface structure and cleanliness were checked by STM imaging. Powder samples of \textbf{1} and \textbf{2} were contained in quartz crucibles, and sublimed from a home-built evaporator at 473 K (\textbf{1}) and 503 K (\textbf{2}) onto the single crystal surfaces held at room temperature.\\
STM and AFM measurements were performed with two instruments from Scienta Omicron operated at a temperature of 5 K and base pressures below 5×10\textsuperscript{-11} mbar. Bias voltages were applied to the sample with respect to the tip. AFM measurements were performed in non-contact mode with a qPlus sensor.~\cite{Giessibl} The sensor was operated in frequency-modulation
mode~\cite{Albrecht} with a constant oscillation amplitude of 0.5 Å. STM images were acquired in constant-current mode, and AFM images were acquired in constant-height mode with CO-functionalized tips at $V = 5$~mV. d\textit{I}/d\textit{V} spectra were acquired in constant-height mode, and d\textit{I}/d\textit{V} maps were acquired in both constant-height and constant-current mode using lock-in amplifiers (SR830 from Stanford Research Systems operating at a frequency of 860 Hz, and HF2Li from Zurich Instruments operating at 691 Hz). STM and AFM images, and d\textit{I}/d\textit{V} spectra and maps, were postprocessed using Gaussian low-pass filters in Wavemetrics Igor Pro software.

\subsection{Density functional theory calculations}

DFT calculations were executed using the AiiDAlab~\cite{yakutovich2021aiidalab,yakutovich2026} applications, based on AiiDA~\cite{pizzi2016aiida} work chains designed for the DFT codes CP2K~\cite{hutter2014cp2k} (systems adsorbed on gold) and Quantum Espresso (band structure calculations).~\cite{giannozzi2009quantum} Surface-adsorbate setups were modeled within a periodic slab scheme. The simulation cell included four gold atomic planes along the [111] orientation. Hydrogen atoms passivated one face of the slab to suppress the Au(111) surface state. A 40 \AA~vacuum layer was included to isolate the system from its periodic images along the axis orthogonal to the slab. Electronic wave functions were represented via the TZV2P Gaussians basis sets  for carbon, nitrogen and hydrogen atoms, and the DZVP basis set for gold atoms. Plane-waves basis set cutoff for the charge density was set at 600 Ry. Norm-conserving Goedecker–Teter–Hutter pseudopotentials were employed. The PBE GGA~\cite{perdew1996generalized}  approximation for the exchange correlation functional was used and Grimme’s D3~\cite{grimme2010consistent}  van der Waals corrections were included. Gold supercells varied in size depending on the adsorbate, ranging from 28.12 \AA~× 26.54 \AA~(corresponding to 598 gold atoms) to 53.63 \AA~× 32.43 \AA~(corresponding to 1534 gold atoms). Geometry optimizations were performed with the bottom two atomic planes constrained while relaxing others, until forces were below 0.005~eV~\AA \textsuperscript{-1}. For AFM simulations, DFT equilibrium geometries and electrostatic potentials were used alongside the probe-particle model.~\cite{hapala2014mechanism}

\smallskip

For the band-structure calculations, ultrasoft pseudopotentials from the SSSP~\cite{hapala2014mechanism}  were employed to model the ionic potentials. A cutoff of 50 Ry (400 Ry) was used for the plane-wave expansion of the wave functions (charge density). The simulation cell contained 15 \AA~of vacuum in the non-periodic directions to minimize interactions among periodic replicas of the system. The thickness of the vacuum region, the sampling of the Brillouin zone and the cutoff ensure convergency of the computed band structures. The atomic positions of the polymer and GNR atoms, and the cell dimension along the ribbon axis were optimized till forces were lower than 0.002 eV~\AA \textsuperscript{-1} and the pressure in the cell was negligible.

\section{Supporting Information}

Solution synthesis and characterization of precursor molecules \textbf{1} and \textbf{2}, scanning probe measurements of precursor, intermediate polymer phases, \textit{poly}-gN-HBC and GNR-gN-HBC, and DFT calculations.

\section{Acknowledgements}\label{acknowledge}
 
This research was supported by the Swiss National Science Foundation under grant Nos. 200020-212875, TMPFP2-210093, 200020-187617, CRSII5 205987 (SNF-PiMag), and the NCCR MARVEL, a National Centre of Competence in Research funded by the Swiss National Science Foundation under grant No. 205602. This research was also supported by a grant from the Swiss National Supercomputing Centre (CSCS) under project ID lp 83.  We acknowledge PRACE for awarding access to the Fenix Infrastructure resources at CSCS, which are partially funded by the European Union’s Horizon 2020 research and innovation program through the ICEI project under the grant agreement No. 800858. We greatly appreciate financial support from the Werner Siemens Foundation (CarboQuant). F.X. thanks the Deutsche Forschungsgemeinschaft (DFG) for a Walter-Benjamin Fellowship (project No. 452269487) and the SNSF Swiss Postdoctoral Fellowship (grant No. 210093). A. N. thanks the financial support by Okinawa Institute of Science and Technology Graduate University, JSPS KAKENHI Grant JP25H01254, and the Max Planck Society.

\section{Author contributions}

S.M., K.M., P.R., A.N. and R.F. conceived and supervised the project. N.B. and S.M. performed the on-surface synthesis and scanning probe measurements, and analyzed the data. Z.Z. and X.-Y.W. synthesized and characterized the precursor molecules in solution. F.X. and N.K. assisted with the scanning probe measurements and data analysis. N.B. and C.A.P. performed the DFT calculations. S.M. wrote the manuscript, with contributions from N.B. All authors contributed to discussing the results and revising the manuscript.

\section{Competing interests}

The authors declare no competing interests.

\section{Additional information}

\textbf{Correspondence and requests for materials} should be addressed to Shantanu Mishra, Pascal Ruffieux or Akimitsu Narita.

\clearpage

\section
  {Supporting Information: One-dimensional carbon nanostructures with periodic graphitic nitrogen substitution}

\setcounter{figure}{0}
\renewcommand{\thefigure}{S\arabic{figure}}

\section{Solution-phase synthesis and characterization}

All reactions working with air- or moisture-sensitive compounds were carried out under an argon atmosphere using standard Schlenk line techniques. 5,10-Dibromo-1,3-diphenyl-2H-cyclopenta[\textit{l}]phenanthren-2-one (\textbf{S6}) was prepared by adapting previously reported synthetic procedures.~\cite{saleh2010triphenylene} Unless otherwise noted, all starting materials and other chemicals were purchased from commercial sources and used without further purification. Thin layer chromatography (TLC) was done on silica gel coated aluminum sheets with F254 indicator and column chromatography separation was performed with silica gel (particle size 0.040-0.063 mm). Nuclear Magnetic Resonance (NMR) spectra were recorded using JEOL JNM-ECA 600 MHz NMR spectrometers. Chemical shifts ($\delta$) were expressed in parts per million (ppm) relative to the residual solvent (CDCl$_3$, $^1$H: 7.26 ppm, $^{13}$C: 77.16 ppm). Coupling constants (\textit{J}) were recorded in Hertz. High-resolution mass spectra (HRMS) were recorded on a Bruker UltrafleXtreme spectrometer by matrix-assisted laser decomposition/ionization (MALDI) using \textit{trans}-2-[3-(4-\textit{tert}-butylphenyl)-2-methyl-2-propenylidene]malononitrile (DCTB) as matrix and calibrating with fleXstandard polymers.

\clearpage

2-(Phenylethynyl)pyridine (\textbf{S3})
\begin{figure}[!h]
\begin{center}
\includegraphics[width=0.9 \linewidth]{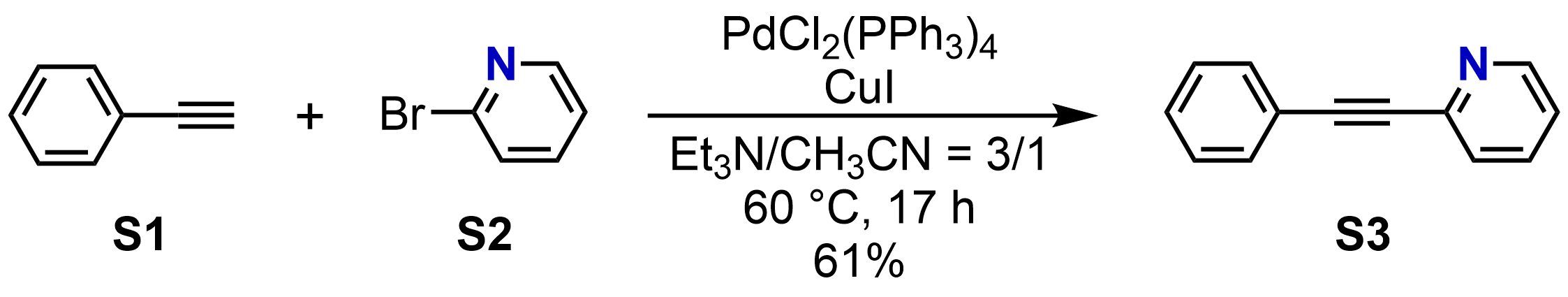}
\caption{Synthesis of compound \textbf{S3}}
\label{s3} 
\end{center}
\end{figure}

In a 100-mL two-neck flask, a solution of ethynylbenzene (\textbf{S1}) (3.4 mL, 28 mmol) and 2-bromopyridine (\textbf{S2}) (3.0 mL, 31 mmol) in triethylamine (30 mL) and acetonitrile (10 mL) was prepared and purged with argon by bubbling for 30 min. CuI (0.61 g, 3.2 mmol) and PdCl$_2$(PPh$_3$)$_4$ (2.2 g, 3.2 mmol) were then added, and the reaction mixture was stirred at 60 °C using a sea sand bath for 17 h under argon atmosphere. After cooling down to room temperature, the mixture was extracted with dichloromethane (100 mL $\times$ 3) and washed once with brine (100 mL $\times$ 1). The organic layer was dried over MgSO$_4$, filtered, and concentrated under reduced pressure. The crude product was purified by column chromatography on silica gel (eluent: hexane/ethyl acetate = 4/1), and the removal of the solvents under reduced pressure afforded compound S3 as a light-yellow solid (3.0 g, 17 mmol, 61\%). The $^1$H NMR spectrum is shown in Figure~S1, which is consistent with a previous report.~\cite{gholinejad2016palladium}

$^1$H NMR (600 MHz, CDCl$_3$, 298 K): $\delta$ 8.63 (s, 1H), 7.68 (td, \textit{J$_1$} = 7.7 Hz, \textit{J$_2$} = 0.9 Hz, 1H), 7.62–7.58 (br, 2H), 7.53 (d, \textit{J} = 7.8 Hz, 1H), 7.39–7.34 (br, 3H), 7.24 (ddd, \textit{J$_1$} = 7.6 Hz, \textit{J$_2$} = 4.9 Hz, \textit{J$_3$} = 1.0 Hz, 1H).

\clearpage

1-(Pyridin-2-yl)-2,4,5-tris(phenyl)-3,6-bis(4-bromophenyl)benzene (\textbf{1})

\begin{figure}[!h]
\begin{center}
\includegraphics[width=0.9 \linewidth]{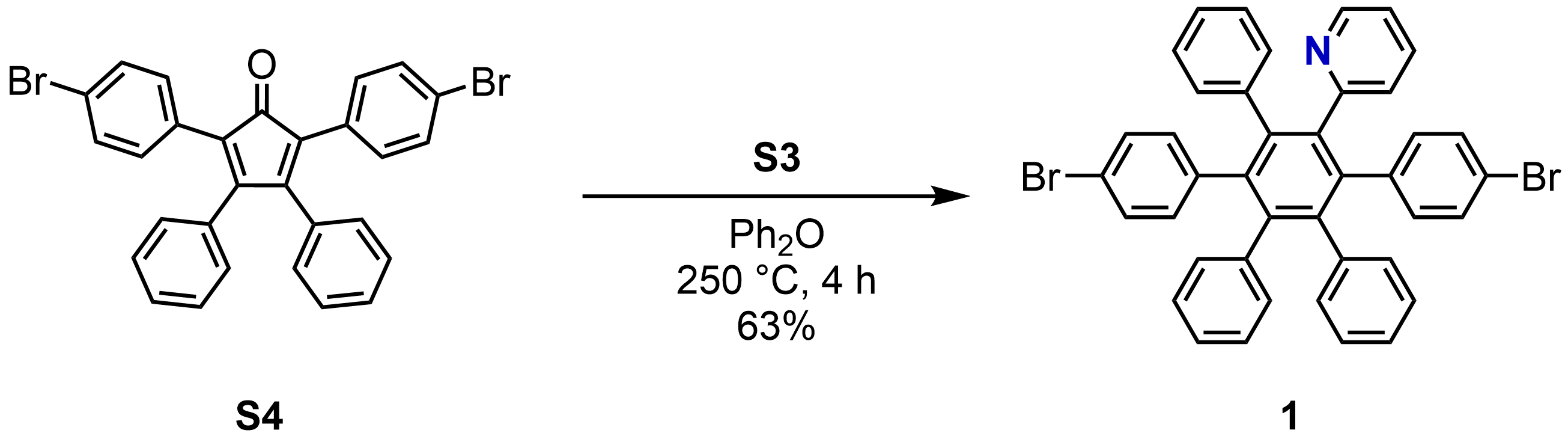}
\caption{Synthesis of compound \textbf{1}}
\label{M1} 
\end{center}
\end{figure}

In a 50-mL two-neck flask, a solution of compound \textbf{S3} (360.8 mg, 2.013 mmol) and 2,5-bis(4-bromophenyl)-3,4-diphenylcyclopenta-2,4-dienone (\textbf{S4}) (1.006 g, 1.855 mmol) in diphenyl ether (10 mL) was placed. The reaction vessel was purged with argon and vacuum three times to create an inert atmosphere, and then the mixture was stirred at 250 °C for 4 h using a sand bath under an argon atmosphere. After the solution was cooled to room temperature, the crude mixture was purified by column chromatography on silica gel (eluent: dichloromethane) to afford compound \textbf{1} as a white solid (814.0 mg, 1.174 mmol, 63\%).

$^1$H NMR (600 MHz, CDCl$_3$, 298 K): $\delta$ 8.20 (ddd, \textit{J$_1$} = 4.9 Hz, \textit{J$_2$} = 1.8 Hz, \textit{J$_3$} = 1.0 Hz, 1H), 7.21 (td, \textit{J$_1$} = 7.7 Hz, \textit{J$_2$} = 0.9 Hz, 1H), 6.98 (dd, \textit{J$_1$} = 7.4 Hz, \textit{J$_2$} = 1.4 Hz, 4H), 6.88–6.66 (br, 21H).

$^{13}$C\{$^1$H\} NMR (151 MHz, CDCl$_3$, 298 K): $\delta$ 159.12, 148.36, 141.45, 140.67, 140.21, 140.05, 139.87, 139.85, 139.73, 139.58, 139.31, 139.20, 134.79, 133.23, 133.13, 133.10, 133.07, 133.03, 132.98, 132.95, 132.33, 132.18, 131.93, 131.44, 131.34, 131.32, 131.24, 130.83, 130.73, 130.06, 130.02, 129.99, 129.95, 127.05, 126.99, 126.51, 125.83, 125.77, 120.72, 119.85, 119.78.

HRMS (MALDI–TOF, positive) (m/z): [M+H]$^+$ calcd for C$_{41}$H$_{28}$Br$_2$N$^+$, 692.0583; found, 692.0556.

\clearpage

5,10-Dibromo-1,3-diphenyl-2H-cyclopenta[\textit{l}]phenanthren-2-one (\textbf{S6})\cite{gholinejad2016palladium}

\begin{figure}[!h]
\begin{center}
\includegraphics[width=0.9 \linewidth]{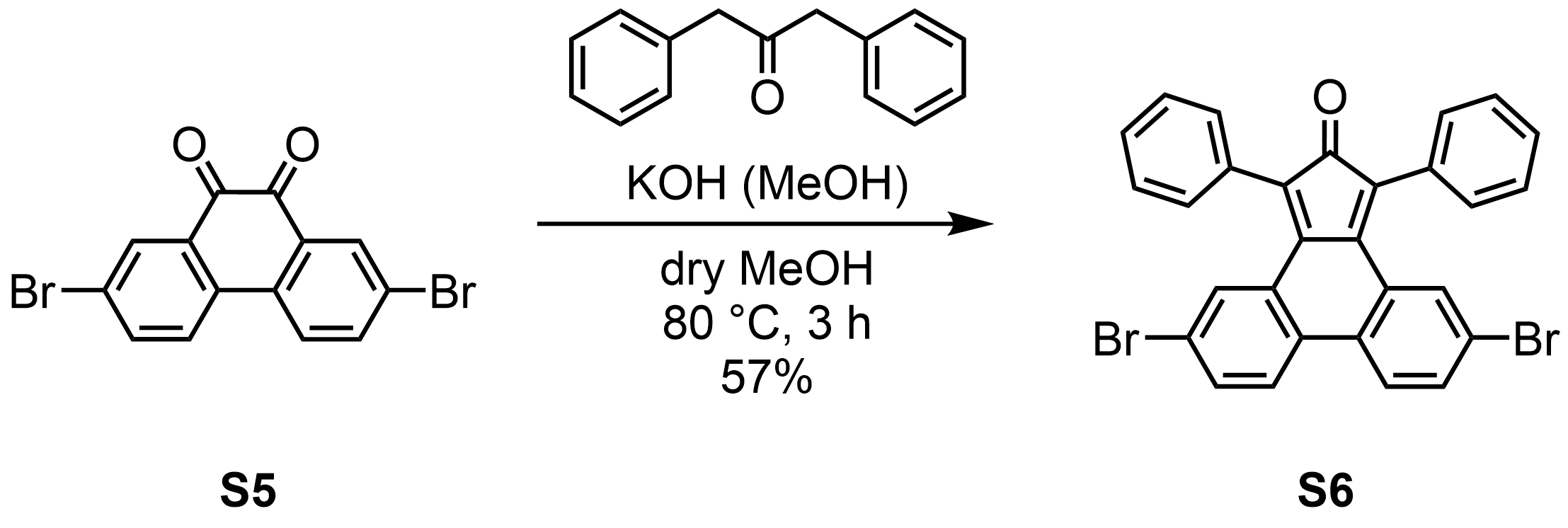}
\caption{Synthesis of compound \textbf{S6}}
\label{s6} 
\end{center}
\end{figure}

In a 100-mL two-neck flask, a solution of 2,7-dibromophenanthrene-9,10-dione (\textbf{S5}) (2.671 g, 7.298 mmol) and 1,3-diphenyl-2-propanone (1.892 g, 8.998 mmol) in dry methanol (15 mL) was placed. The reaction mixture was stirred at 80 °C using a sea sand bath under argon atmosphere. A solution of KOH (431.6 mg, 7.692 mmol) in dry methanol (25 mL) was added dropwise over 1 h via a dropping funnel, and the resulting mixture was stirred for an additional 2 h at the same temperature. After cooling down to room temperature, the mixture was filtered, and the collected solid was washed with toluene (20 mL $\times$ 5) to afford compound \textbf{S2} as a deep green solid (2.263 g, 4.189 mmol, 57\%). The $^1$H NMR spectrum is shown in Figure~S4, which is consistent with a previous report.~\cite{saleh2010triphenylene}

$^1$H NMR (600 MHz, CDCl$_3$, 298 K): $\delta$ 7.66 (d, \textit{J} = 2.0 Hz, 2H), 7.61 (d, \textit{J} = 8.7 Hz, 2H), 7.48–7.36 (br, 12H).

\clearpage

10-(Pyridin-2-yl)-9,11,12-tris(phenyl)-2,7-dibromotriphrnylene (\textbf{2})

\begin{figure}[!h]
\begin{center}
\includegraphics[width=0.9 \linewidth]{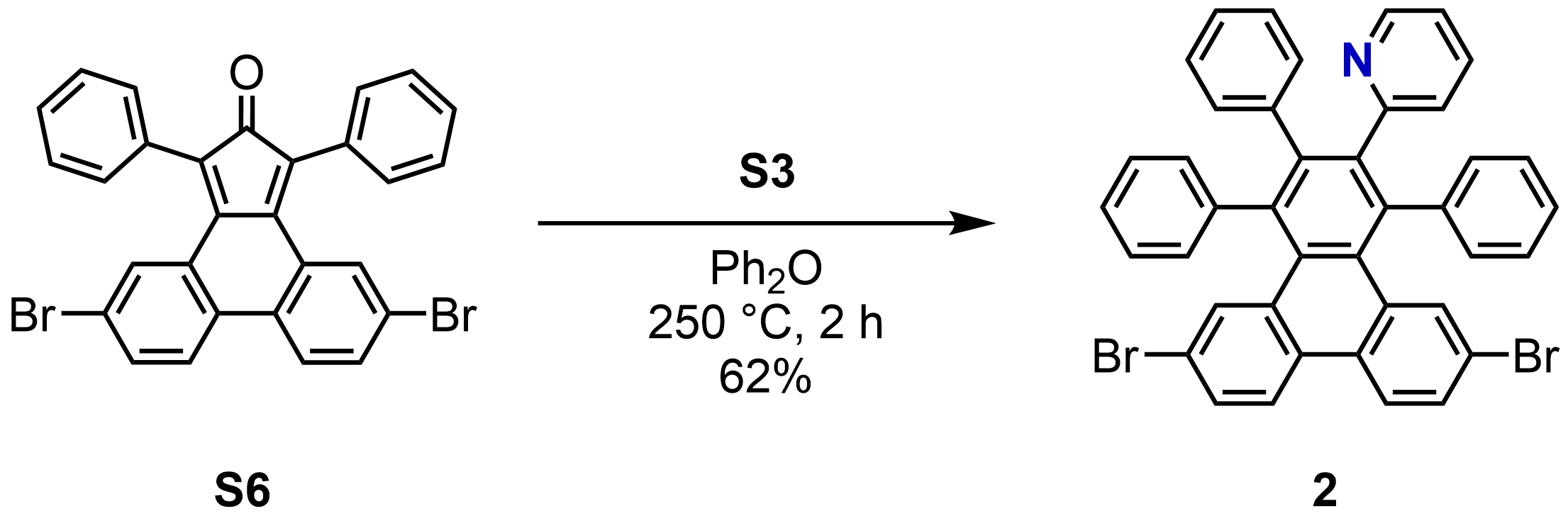}
\caption{Synthesis of compound \textbf{2}}
\label{m2} 
\end{center}
\end{figure}

In a 50-mL two-neck flask, a solution of compound \textbf{S3} (378.4 mg, 2.111 mmol) and compound \textbf{S6} (1.016 g, 1.881 mmol) in diphenyl ether (10 mL) was placed. The reaction vessel was purged with argon and vacuum three times to create an inert atmosphere, and then the mixture was stirred at 250 °C for 2 h using a sand bath under an argon atmosphere. After the solution was cooled to a room temperature, the reaction mixture was purified by column chromatography on silica gel (eluent: dichloromethane) to afford compound \textbf{2} as a slightly yellowish white solid (804.1 mg, 1.163 mmol, 62\%).

$^1$H NMR (600 MHz, CDCl$_3$, 298 K): $\delta$ 8.22–8.20 (br, 3H). 7.69 (d, \textit{J} = 2.0 Hz, 1H), 7.67 (d, \textit{J} = 2.0 Hz, 1H), 7.49 (ddd, \textit{J$_1$} = 8.7 Hz, \textit{J$_2$} = 1.7 Hz, \textit{J$_3$} = 1.7 Hz, 2H), 7.24 (td, \textit{J$_1$} = 7.7 Hz, \textit{J$_2$} = 0.9 Hz, 1H), 7.20–7.14 (br, 7H), 7.06–7.04 (br, 3H), 6.94–6.88 (br, 3H), 6.84 (ddd, \textit{J$_1$} = 7.6 Hz, \textit{J$_2$} = 4.9 Hz, \textit{J$_3$} = 1.1 Hz, 1H), 6.80–6.78 (br, 2H), 6.69–6.68 (br, 1H).

$^{13}$C\{$^1$H\} NMR (151 MHz, CDCl$_3$, 298 K): $\delta$ 159.15, 148.28, 141.92, 141.80, 141.74, 140.90, 140.56, 139.75, 137.96, 137.61, 134.55, 134.50, 132.98, 132.34, 132.31, 132.03, 131.98, 131.81, 131.55, 131.07, 131.00, 130.72, 130.35, 129.83, 129.77, 129.74, 129.70, 128.70, 128.59, 128.48, 128.41, 127.09, 127.05, 127.02, 126.99, 126.86, 125.70, 124.60, 120.65, 120.27, 120.25.

HRMS (MALDI–TOF, positive) (m/z): [M+H]$^+$ calcd for C$_{41}$H$_{26}$Br$_2$N$^+$, 690.0427; found, 690.0460.

\clearpage

\begin{figure}[!h]
\begin{center}
\includegraphics[width= 0.9\linewidth]{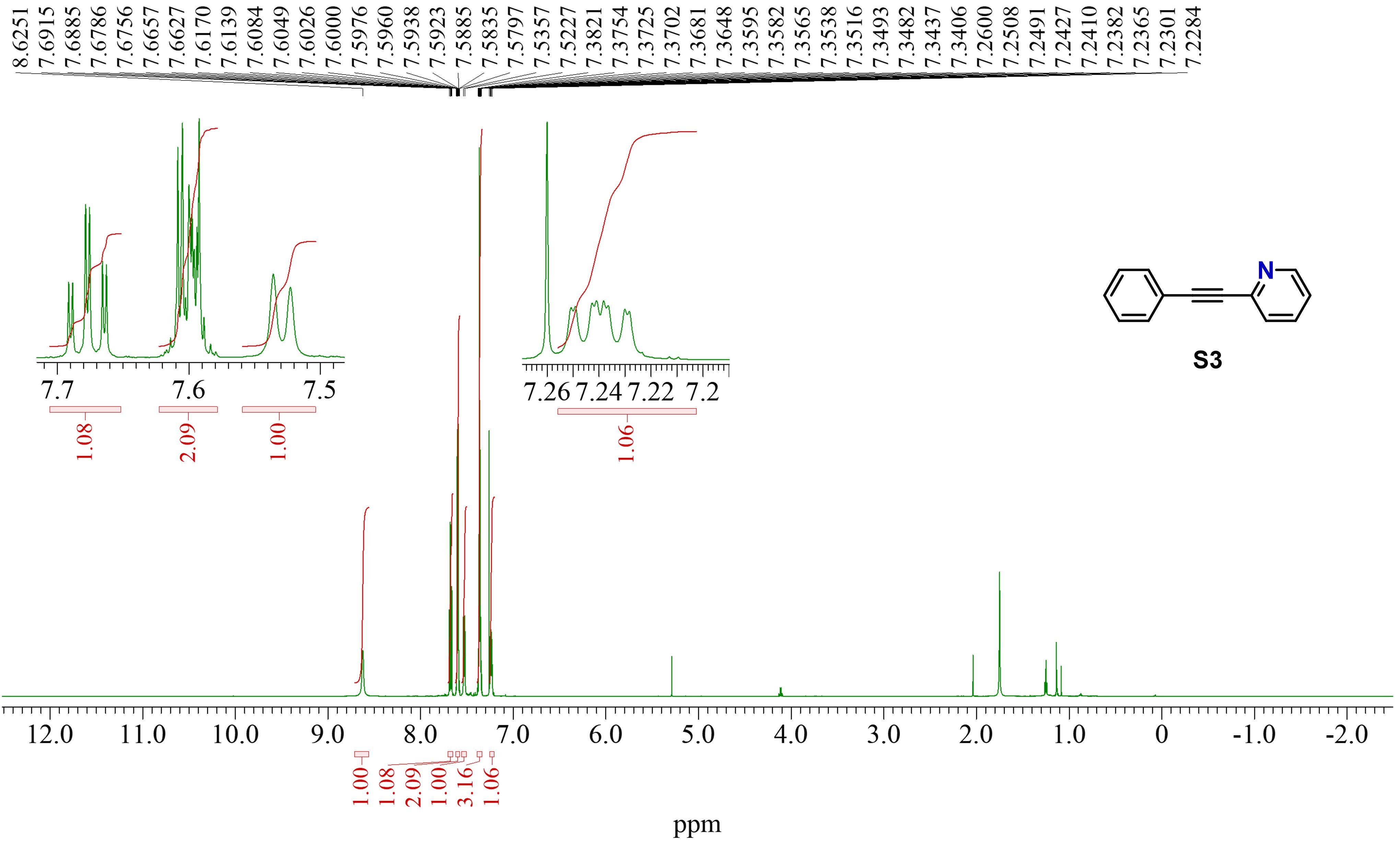}
\caption{$^1$H NMR spectrum of compound \textbf{S3} (CDCl$_3$, 600 MHz, 298 K).}
\label{nmr1} 
\end{center}
\end{figure}

\begin{figure}[!h]
\begin{center}
\includegraphics[width=0.9 \linewidth]{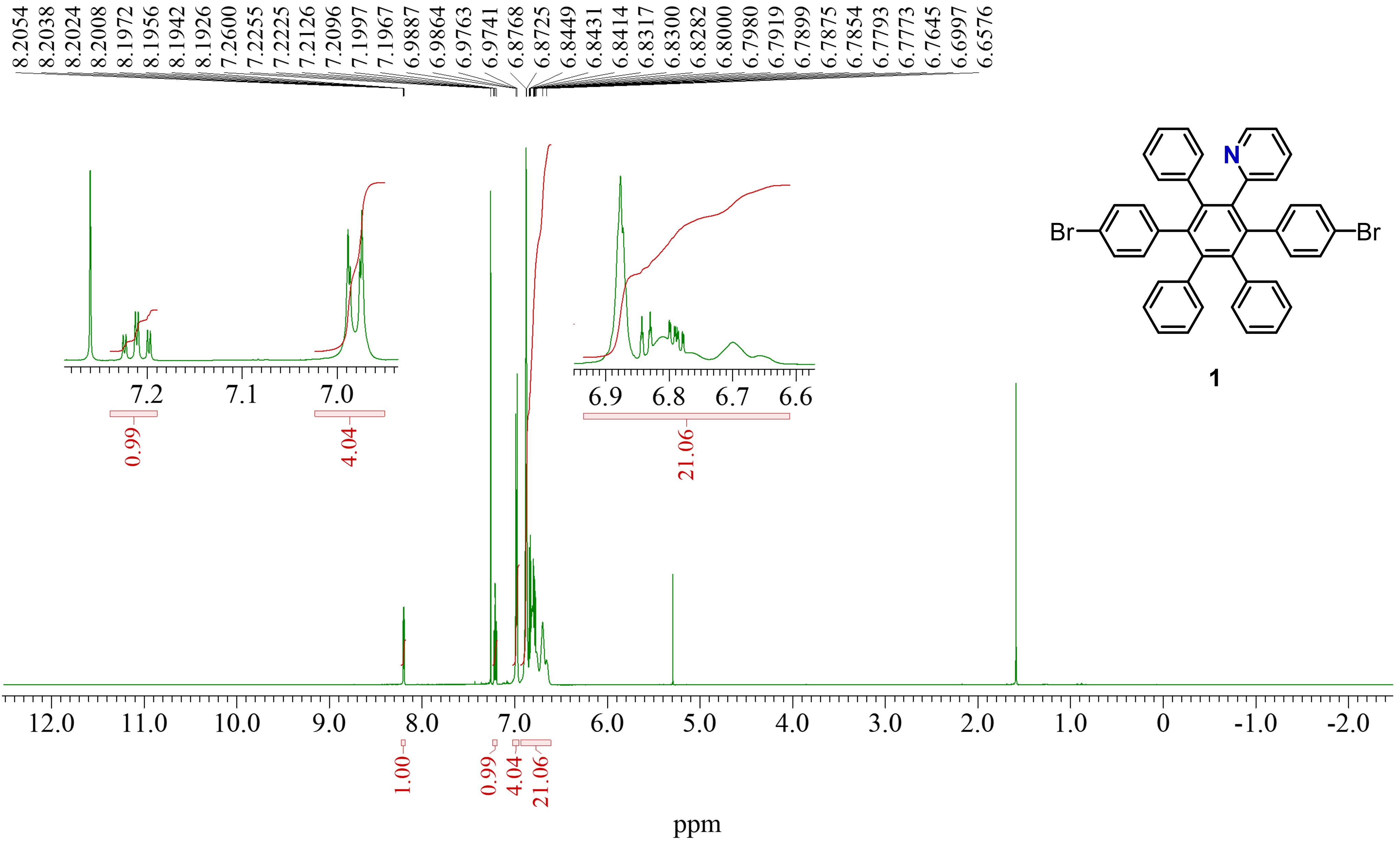}
\caption{$^1$H NMR spectrum of compound \textbf{1} (CDCl$_3$, 600 MHz, 298 K).
}
\label{nmr2} 
\end{center}
\end{figure}

\begin{figure}[!h]
\begin{center}
\includegraphics[width=0.9 \linewidth]{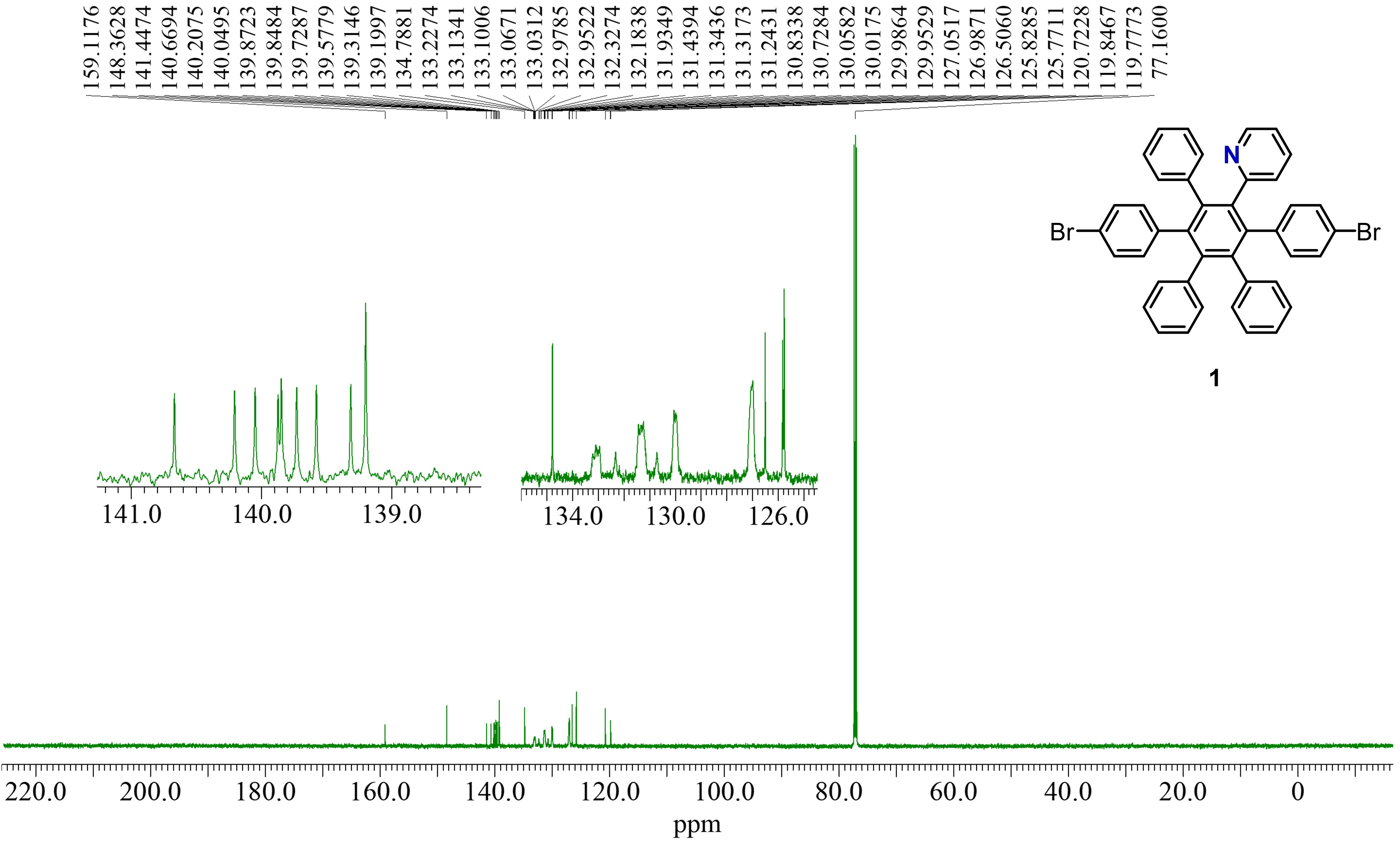}
\caption{$^{13}$C NMR spectrum of compound \textbf{1} (CDCl$_3$, 126 MHz, 298 K).}
\label{nmr3} 
\end{center}
\end{figure}

\begin{figure}[!h]
\begin{center}
\includegraphics[width=0.9 \linewidth]{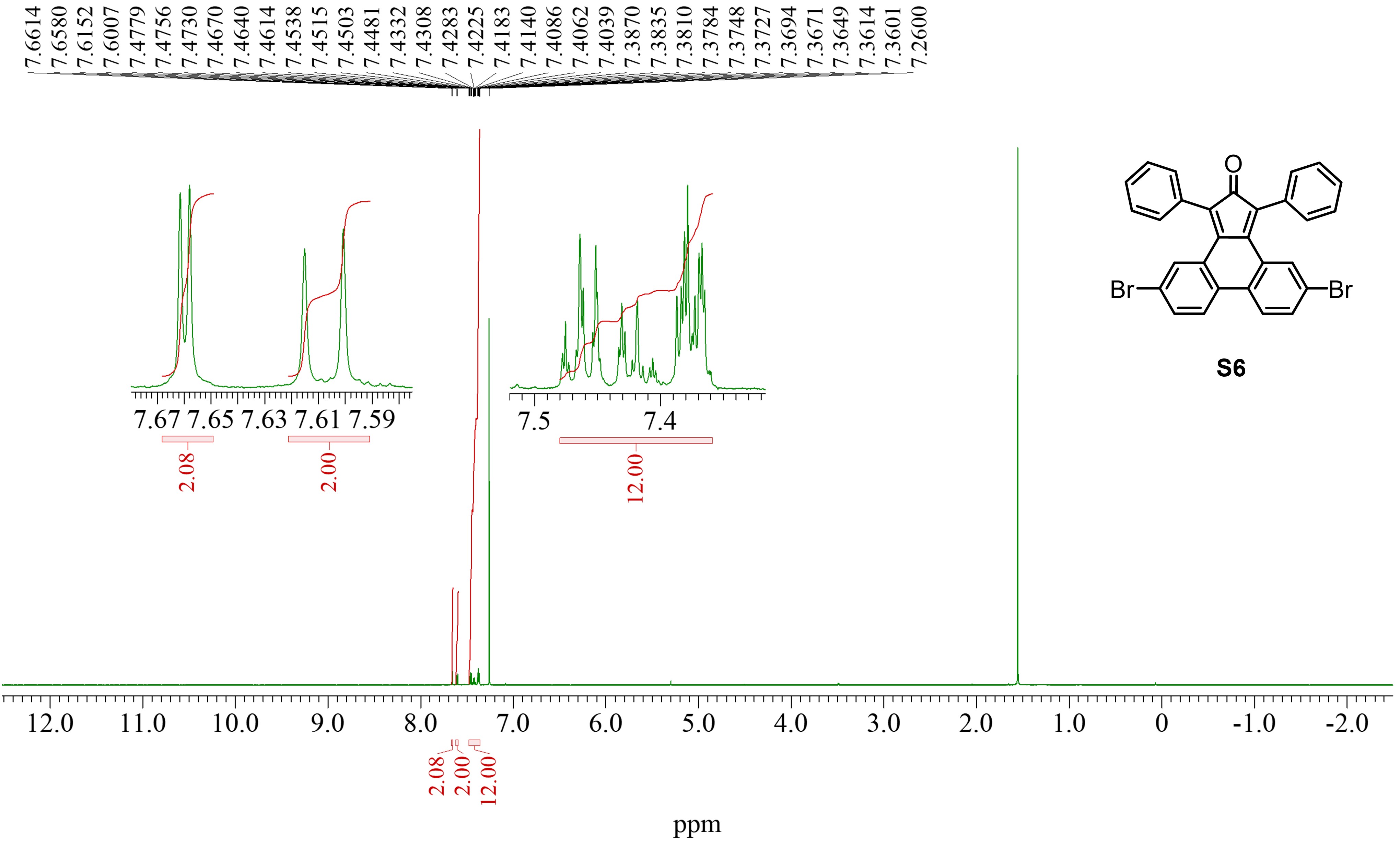}
\caption{$^1$H NMR spectrum of compound \textbf{S6} (CDCl$_3$, 600 MHz, 298 K).}
\label{nmr4} 
\end{center}
\end{figure}

\begin{figure}[!h]
\begin{center}
\includegraphics[width=0.9 \linewidth]{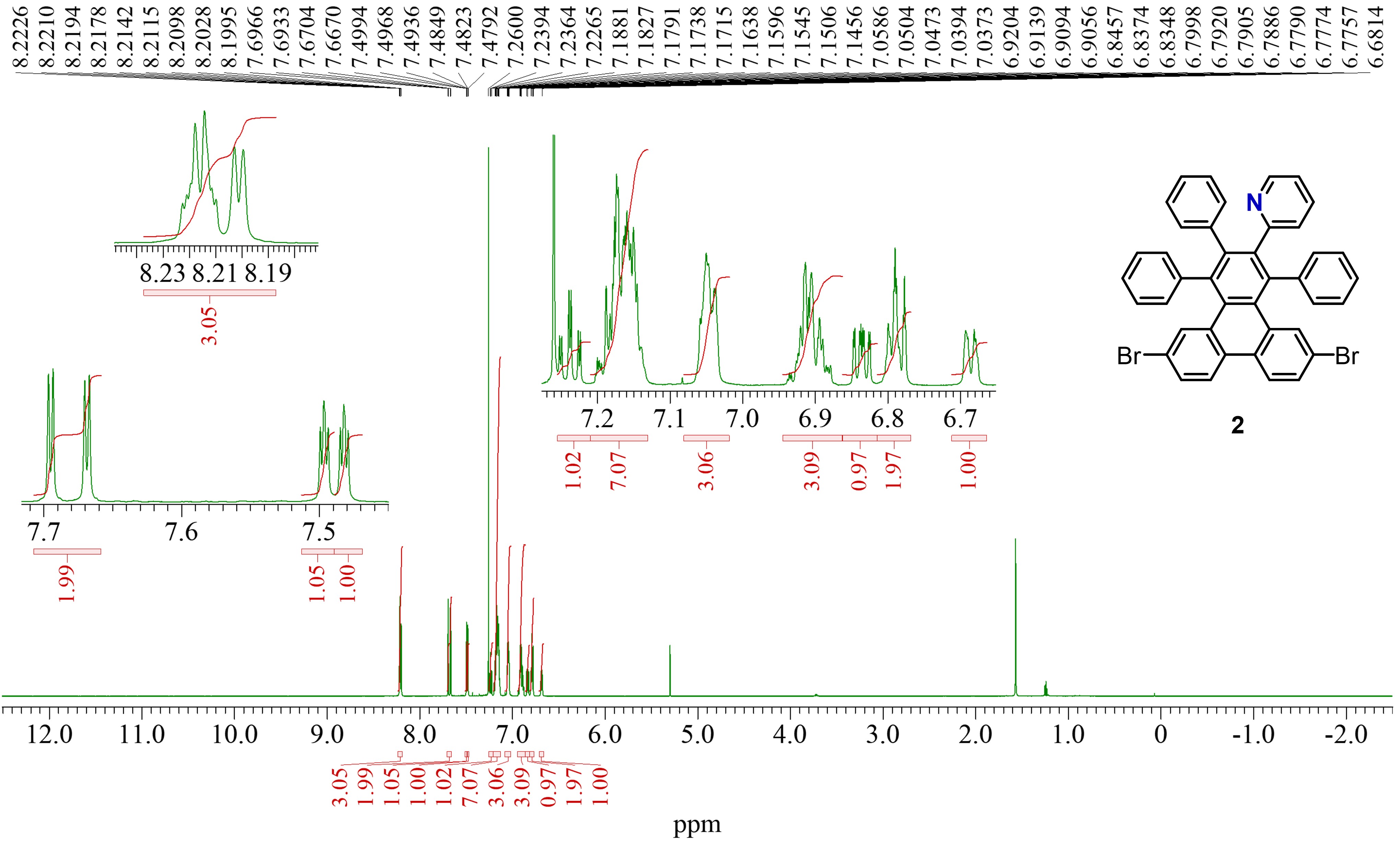}
\caption{$^1$H NMR spectrum of compound \textbf{2} (CDCl$_3$, 600 MHz, 298 K).
}
\label{nmr5} 
\end{center}
\end{figure}

\begin{figure}[!h]
\begin{center}
\includegraphics[width=0.9 \linewidth]{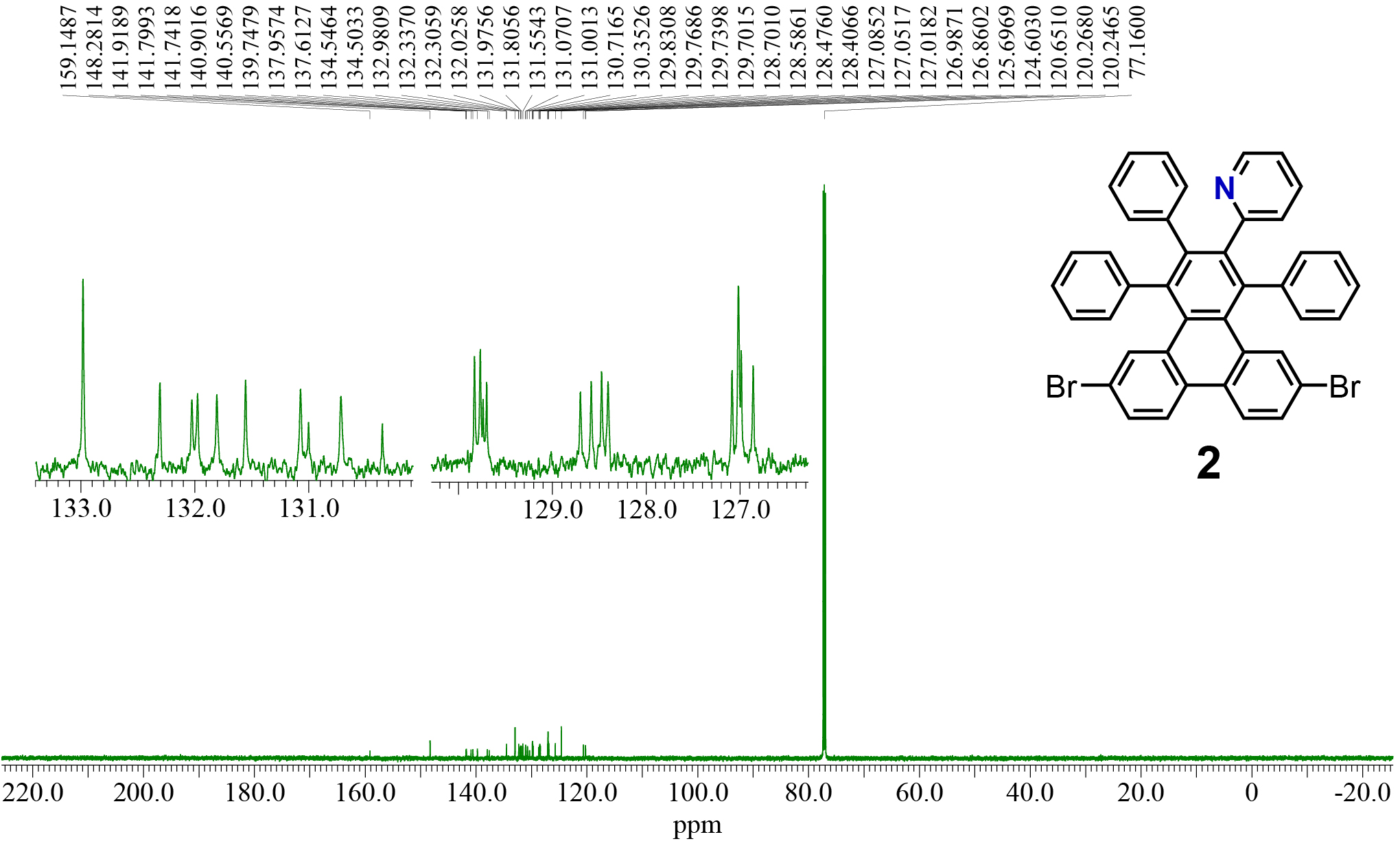}
\caption{$^{13}$C NMR spectrum of compound \textbf{2} (CDCl$_3$, 126 MHz, 298 K).}
\label{nmr6} 
\end{center}
\end{figure}

\clearpage

\section{Scanning probe measurements and calculations}

\begin{figure}[!h]
\begin{center}
\includegraphics[width=0.9 \linewidth]{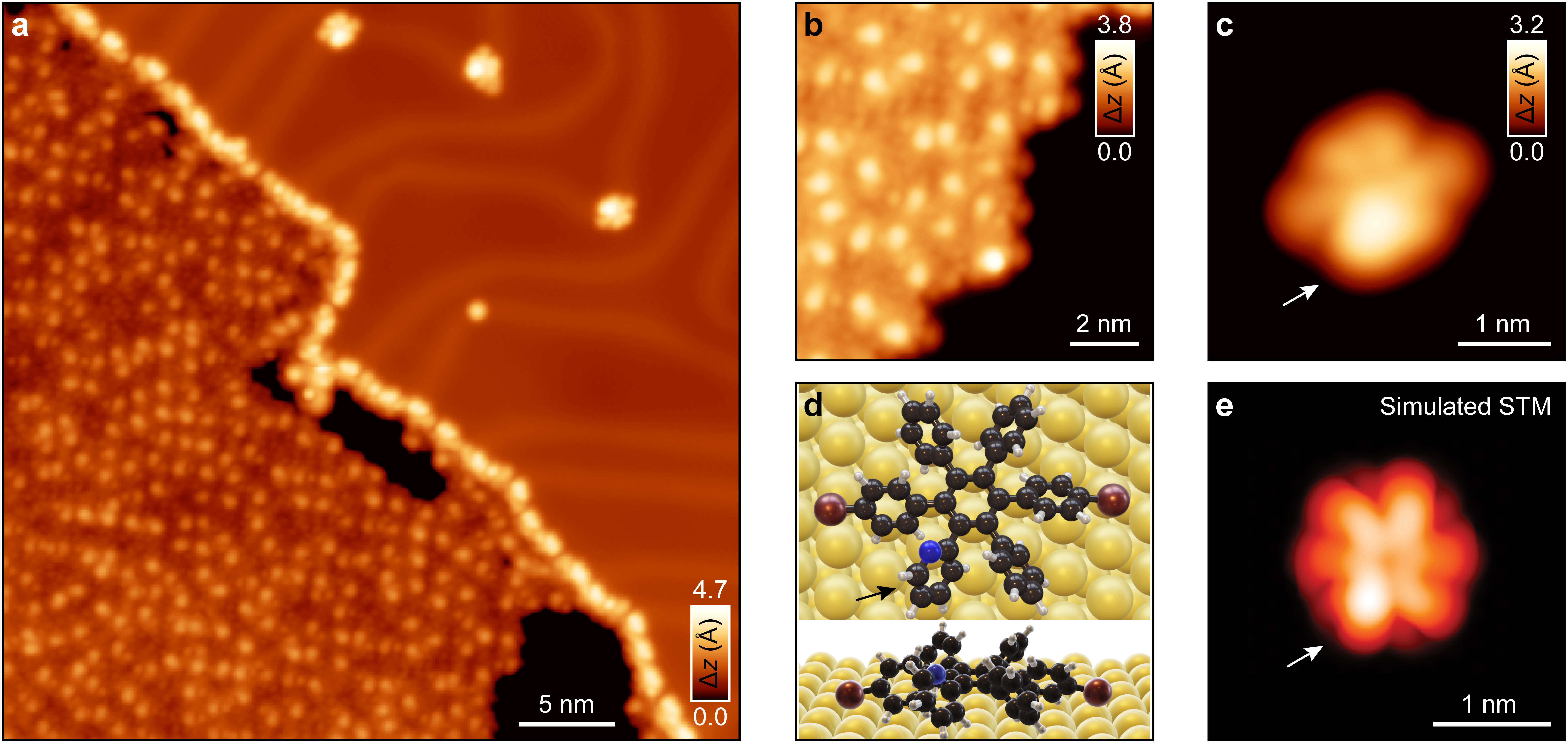}
\caption{Characterization of \textbf{1} on Au(111). (a) STM overview image of the sample after deposition of \textbf{1} on Au(111) held at room temperature ($V = -1.5$~V, $I = 50$~pA). The surface consists of self-assembled molecular islands (lower terrace), and individual molecules anchored at the Au(111) herringbone elbow sites (upper terrace) and along the step edges. (b) High-resolution STM image of a self-assembled island ($V = -0.4$~V, $I = 100$~pA). (c) High-resolution STM image of \textbf{1} ($V = -1.0$~V, $I = 200$~pA). Six lobes corresponding to the outer non-planar rings of the molecule are visible in the STM image. The pyridyl ring, which appears brighter due to the higher density of states at the nitrogen site, is indicated by an arrow. (d) Top and side views of the adsorption geometry of \textbf{1} on Au(111) optimized using DFT. The pyridyl ring is indicated by an arrow. (e) Simulated STM image of \textbf{1} on Au(111) ($V = -1.1$~V, $I = 50$~pA). The enhanced contrast at the pyridyl ring (indicated by an arrow) is in good agreement with the experimental observation in (c).
}
\label{320RT} 
\end{center}
\end{figure}

\begin{figure}[!h]
\begin{center}
\includegraphics[width=0.9 \linewidth]{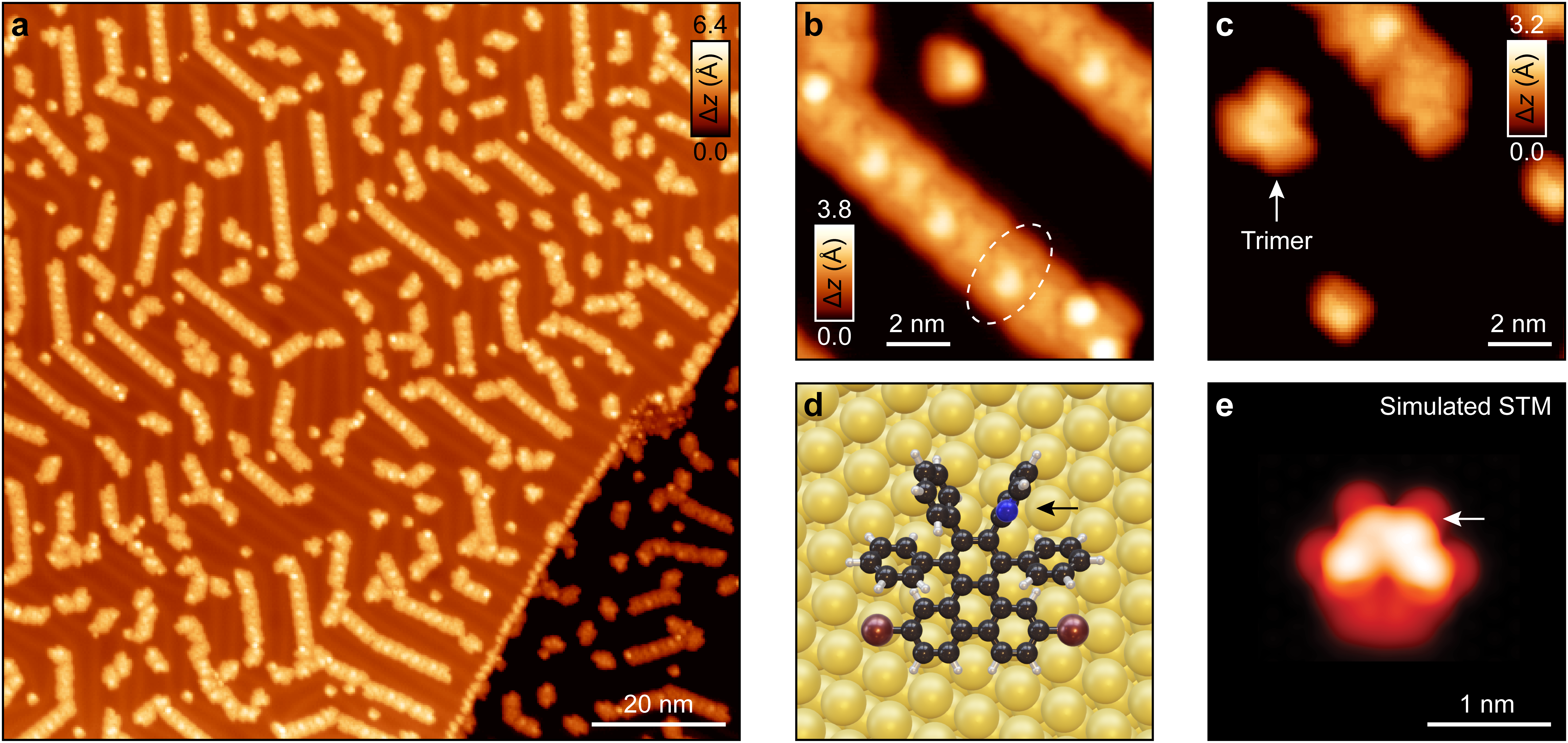}
\caption{Characterization of \textbf{2} on Au(111). (a) STM overview image of the sample after deposition of \textbf{2} on Au(111) held at room temperature ($V = -1.5$~V, $I = 50$~pA). The surface consists of self-assembled chains on the fcc regions of the Au(111) surface, self-assembled clusters, and individual molecules anchored at the Au(111) herringbone elbow sites and along the step edges. (b) High-resolution STM image of self-assembled chains and an individual molecule ($V = -1.5$~V, $I = 50$~pA). The chains consist of repeating dimers of \textbf{2}, with a constituent dimer indicated by the dashed ellipse. Similar to \textbf{1}, the pyridyl ring of \textbf{2} appears brighter in STM imaging. (c) High-resolution STM image of molecular clusters and individual molecules ($V = -1.5$~V, $I = 50$~pA). A self-assembled trimer is indicated by an arrow. (d) Top view of the adsorption geometry of \textbf{2} on Au(111) optimized using DFT. (e) Simulated STM image of \textbf{2} on Au(111) ($V = -0.8$~V, $I = 100$~pA). The pyridyl ring is indicated by arrows in (d, e).}
\label{467RT} 
\end{center}
\end{figure}

\begin{figure}[!h]
\begin{center}
\includegraphics[width=0.9 \linewidth]{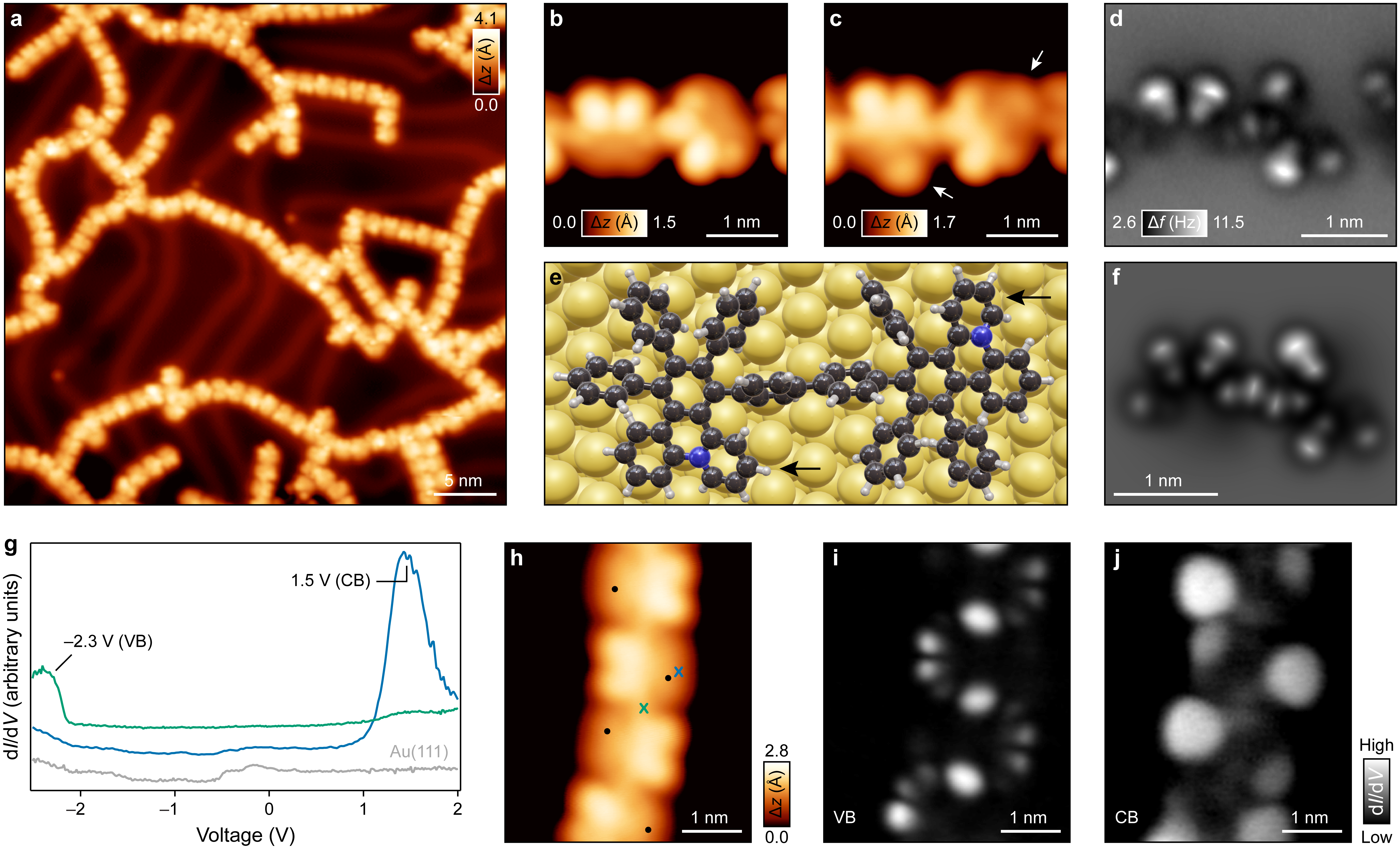}
\caption{Characterization of \textit{poly}-\textbf{1} on Au(111). (a) STM overview image after annealing \textbf{1} on Au(111) at 250~$^\circ$C ($V = -1.2$~V, $I = 50$~pA). Ullmann-like coupling of \textbf{1} is evidenced by the presence of polymer chains on the surface. (b) High-resolution in-gap STM image of a \textit{poly}-\textbf{1} segment ($V = -0.4$~V, $I = 15$~pA). Partial planarization of the individual units has occurred due to C--N bond formation. The non-planar phenyl rings appear as bright lobes. (c) High-resolution STM image of the same segment acquired at $V = 1.5$~V and $I = 15$~pA, revealing prominent lobes at the gN sites (indicated by arrows). (d) AFM image of the segment. The non-planar phenyl rings in each unit appear brighter (that is, with an increased $\Delta f$) due to stronger repulsive forces, whereas the comparatively planar pyrido[1,2-\textit{a}]quinolin-11-ium moieties, which are adsorbed closer to the surface, are not resolved. The left and right units in (b--d) exhibit on- and off-axis C--N bond formation, respectively. STM set-point: $V = 5$~mV and $I = 100$~pA on Au(111); $\Delta z = 4.7$~\AA. (e) Top view of the adsorption geometry of a dimer on Au(111) optimized using DFT and corresponding to the segment in (b--d). The nitrogen-containing six-membered rings are indicated by arrows. (f) Corresponding simulated AFM image of the dimer. (g) d\textit{I}/d\textit{V} spectra acquired on a \textit{poly}-\textbf{1} segment. Open feedback parameters: $V = -2.5$~V and $I = 300$~pA on the individual units; $V_{\mathrm{rms}} = 16$~mV. Acquisition positions are indicated by the colored crosses in the STM image in (h) ($V = -1.0$~V, $I = 130$~pA). Filled circles in (h) denote the positions of the nitrogen atoms. The spectra reveal two resonances at $V = -2.3$~V and $V = 1.5$~V, corresponding to the valence band (VB) and conduction band (CB), respectively. (i, j) Constant-current d\textit{I}/d\textit{V} maps of the segment in (h) acquired at voltages corresponding to the resonances indicated in (g). Scanning parameters: $V = -2.3$~V, $I = 200$~pA (i), and $V = 1.5$~V, $I = 130$~pA (j); $V_{\mathrm{rms}} = 16$~mV. The state at $V = 1.5$~V is predominantly localized at the gN sites. The location of the state at positive bias indicates that it corresponds to an unoccupied nitrogen band. Thus, similar to \textit{poly}-gN-HBC, \textit{poly}-\textbf{1} adopts a cationic state on Au(111).}
\label{320poly1} 
\end{center}
\end{figure}

\begin{figure}[!h]
\begin{center}
\includegraphics[width=0.70 \linewidth]{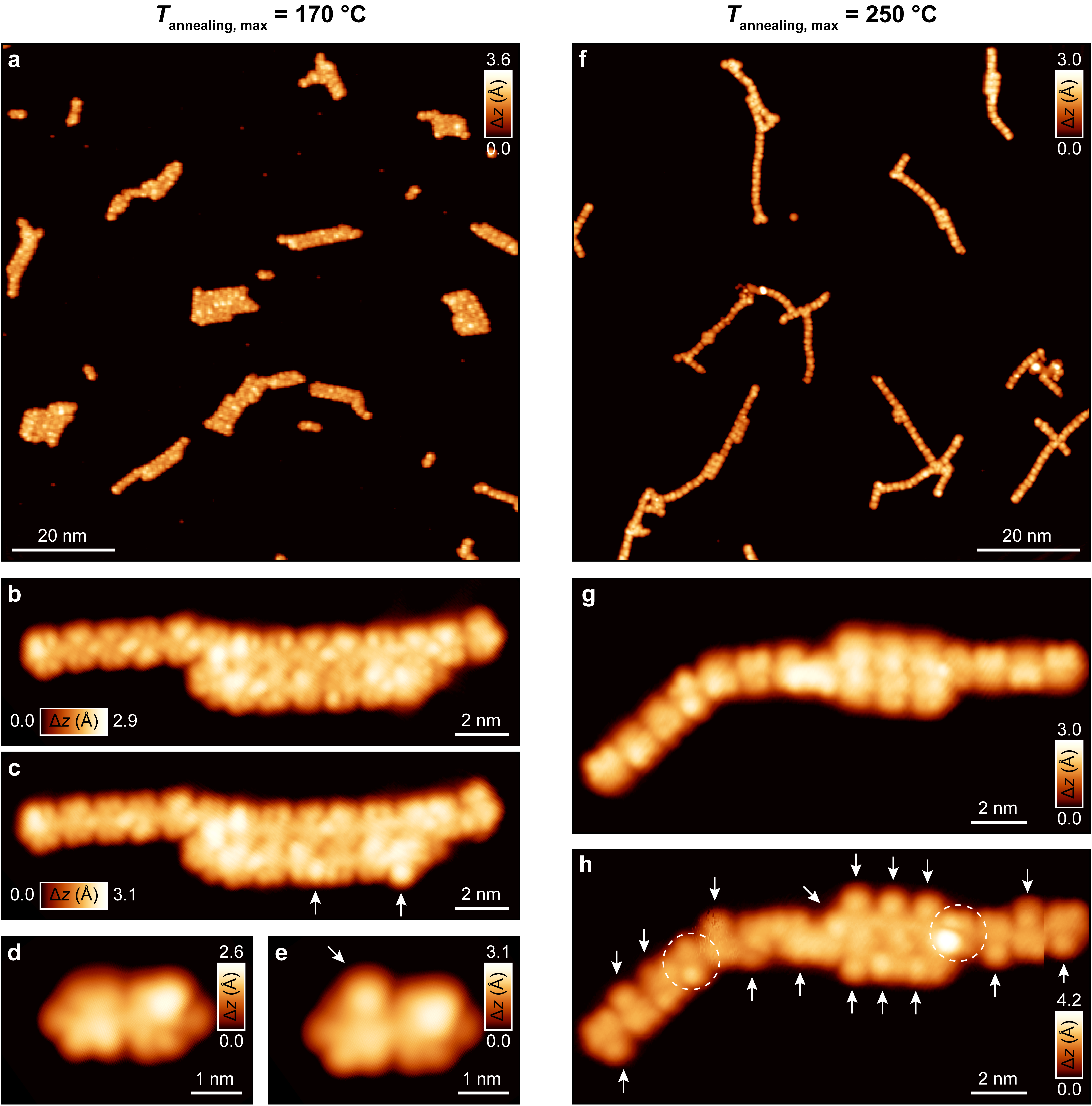}
\caption{C–N bond formation as a function of annealing temperature. (a) STM overview image after annealing \textbf{1} on Au(111) at 170~$^\circ$C for 20~minutes ($V = -1.0$~V, $I = 10$~pA). The surface mostly consists of self-assembled polymer chains and a few isolated short chains. (b) High-resolution in-gap STM image of two interdigitated polymer chains consisting of a total of 21 individual units ($V = -1.0$~V, $I = 60$~pA). (c) High-resolution STM image of the same structure acquired at $V = 1.5$~V and $I = 40$~pA. Only 2 out of the 21 units exhibit the characteristic lobes (indicated by arrows) associated with the gN sites, which is indicative of C--N bond formation. Analysis of multiple STM images revealed that 14 out of 136 units ($\sim$10\%) exhibited C--N bond formation at 170~$^\circ$C. (d) High-resolution in-gap STM image of a dimer ($V = -0.2$~V, $I = 50$~pA), where the left unit exhibits planarization at the upper end, while the right unit shows the characteristic bright appearance of the pyridyl ring, as observed for \textbf{1} (Figure~S11). (e) High-resolution STM image of the same dimer acquired at $V = 1.5$~V and $I = 40$~pA. The left unit exhibits the characteristic lobe associated with a gN site (indicated by an arrow). (f) STM overview image of the same sample after further annealing the sample at 250~$^\circ$C for 15~minutes ($V = -1.0$~V, $I = 15$~pA). In contrast to (a), the polymer chains are no longer self-assembled but dispersed on the surface, indicating a high degree of planarization due to C--N bond formation. (g) High-resolution in-gap STM image of two interdigitated polymer chains consisting of a total of 18 individual units ($V = -1.0$~V, $I = 60$~pA). (h) High-resolution STM image of the same structure acquired at $V = 1.5$~V and $I = 2$~pA. Sixteen out of the 18 units exhibit the characteristic lobes associated with the gN sites (indicated by arrows), while only 2 units (indicated by dashed circles) do not exhibit these lobes. This indicates that C--N bond formation is nearly complete at 250~$^\circ$C.}
\label{320poly1_anne} 
\end{center}
\end{figure}
\clearpage

\begin{figure}[!h]
\begin{center}
\includegraphics[width=0.9 \linewidth]{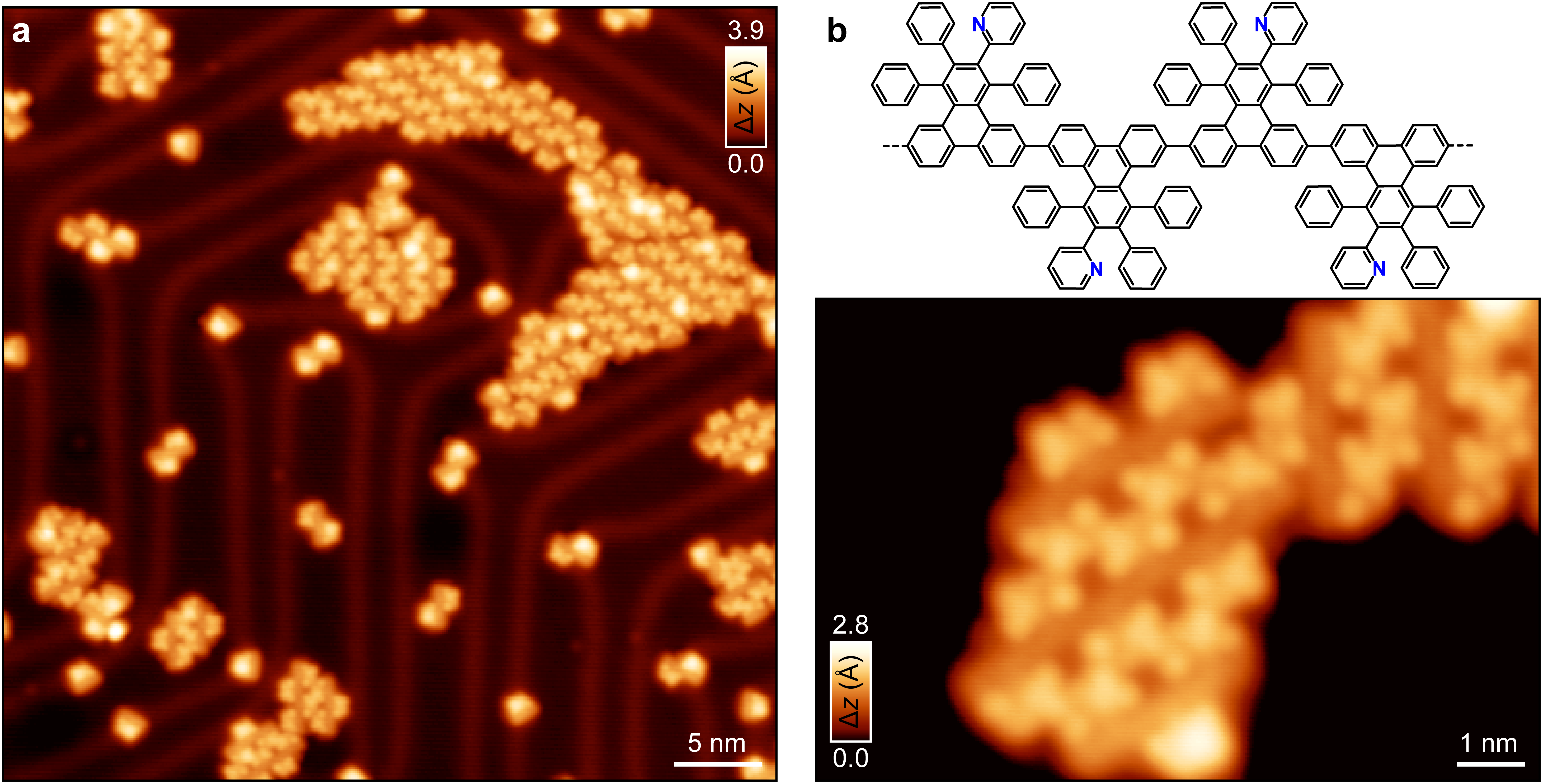}
\caption{Characterization of an intermediate polymer phase after annealing \textbf{2} on Au(111). (a) STM overview image after annealing \textbf{2} on Au(111) at 180~$^\circ$C ($V = -1.0$~V, $I = 50$~pA). The surface consists of self-assembled polymer chains, as well as isolated monomers to tetramers. The high degree of self-assembly indicates a general absence of C--N bond formation, consistent with the results shown in Figure~S14 for a similar annealing temperature. (b) Chemical structure of the intermediate polymer (top) and high-resolution STM image of self-assembled polymer chains (bottom; $V = -0.7$~V, $I = 50$~pA). The central planar part of each polymer chain, flanked by features approximately 0.5~\AA\ higher on either side, indicates that C--N bond formation has not occurred.}
\label{467poly1} 
\end{center}
\end{figure}
\clearpage

\begin{figure}[!h]
\begin{center}
\includegraphics[width=0.9 \linewidth]{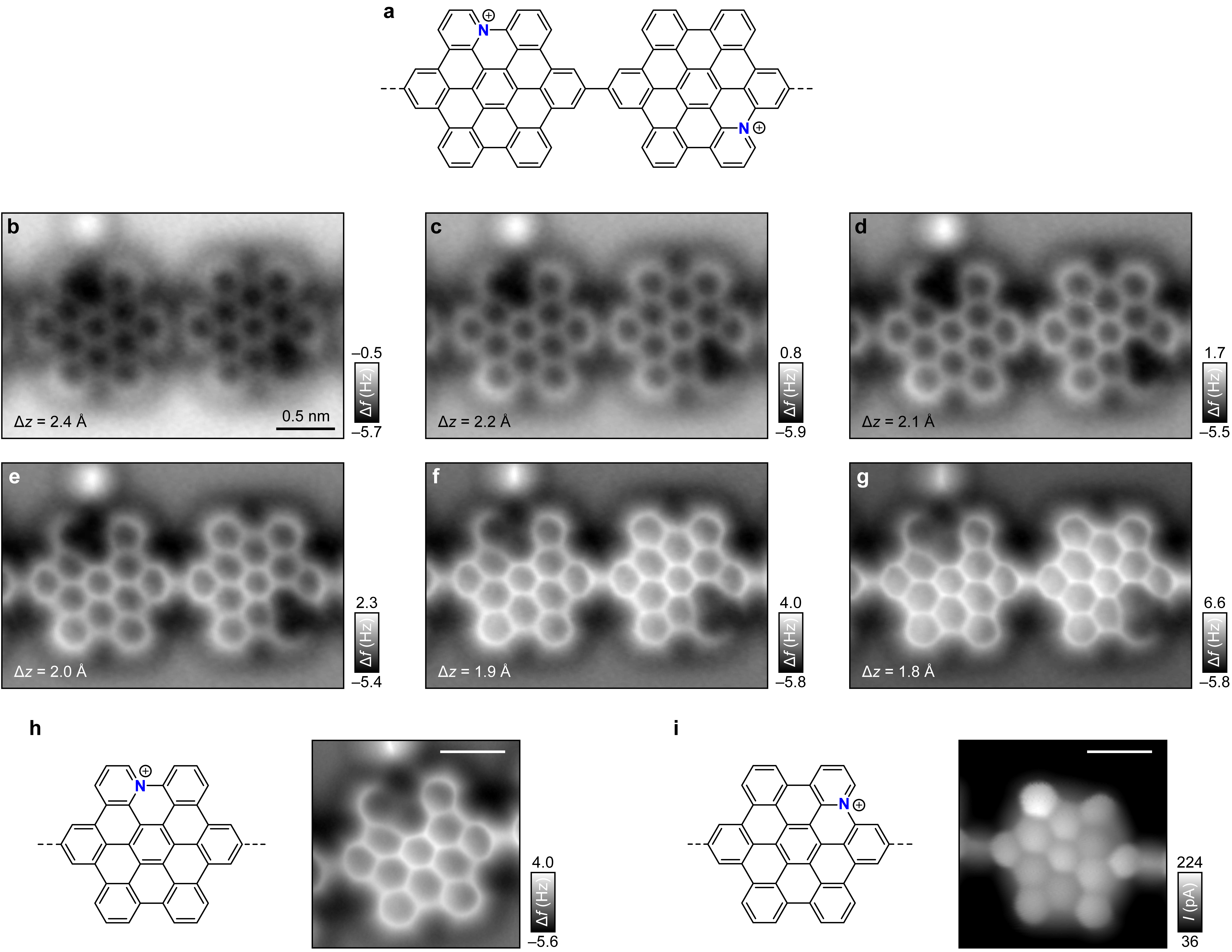}
\caption{Bond-resolved STM and AFM imaging of \textit{poly}-gN-HBC. (a) Chemical structure of a \textit{poly}-gN-HBC segment consisting of two gN-HBC units. (b--g) High-resolution AFM images of the segment in (a) acquired at different $\Delta z$ values, as indicated in the respective panels. STM set-point: $V = 5$~mV and $I = 100$~pA on Au(111). (h) Chemical structure (left) and high-resolution AFM image (right) of a gN-HBC unit. (i) Chemical structure (left) and high-resolution constant-height STM image acquired with a CO-functionalized tip (right) of a gN-HBC unit ($V = 5$~mV). Note the blurred appearance of the nitrogen-containing rings in (i). Scale bars: 0.5 nm.}
\label{320a_AFM} 
\end{center}
\end{figure}
\clearpage

\begin{figure}[!h]
\begin{center}
\includegraphics[width=0.9 \linewidth]{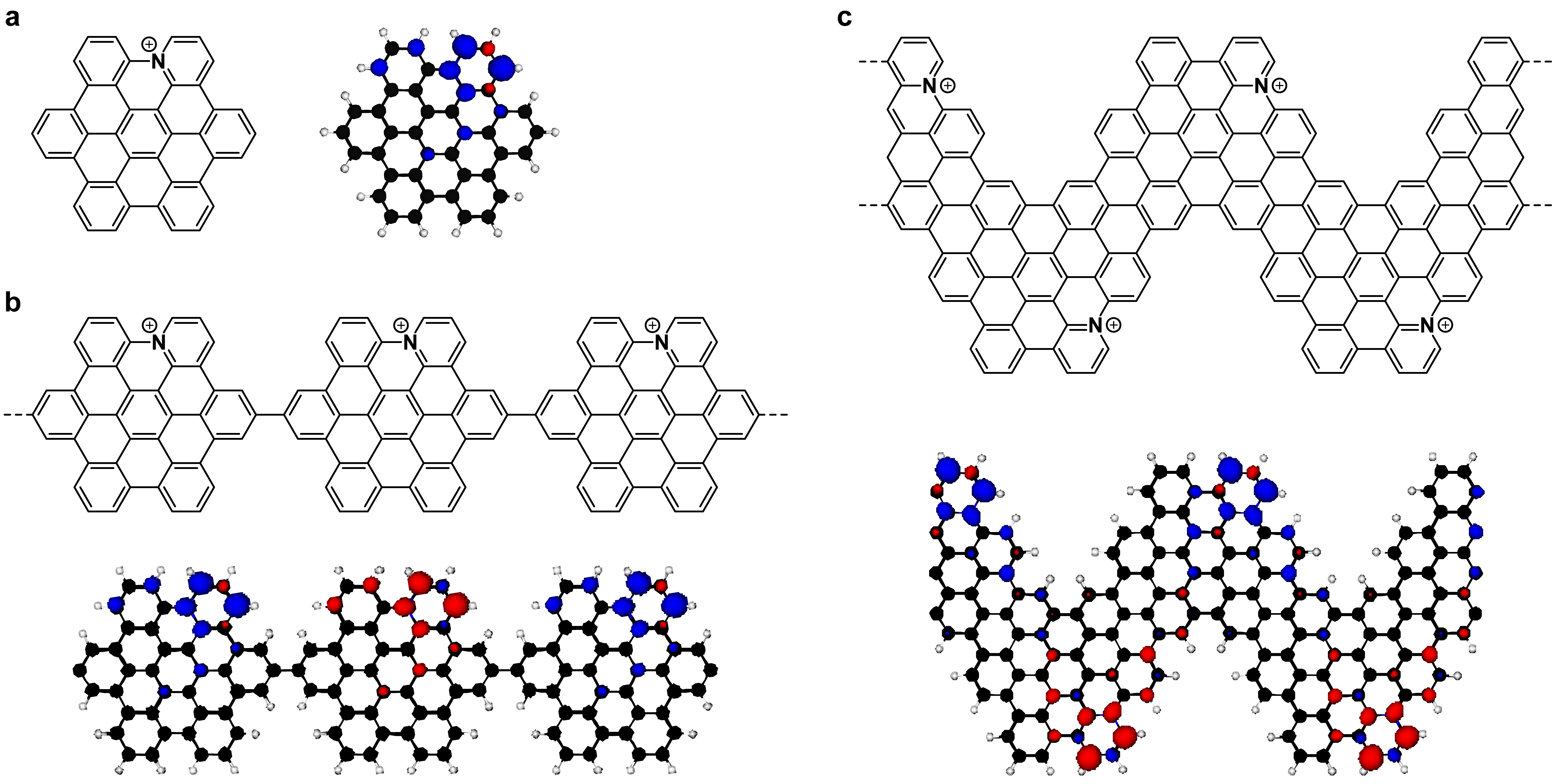}
\caption{Spin-polarized DFT calculations. (a--c) Chemical structures and spin-density isosurface plots (obtained using DFT) of gN-HBC (a), \textit{poly}-gN-HBC in the neutral antiferromagnetic state (b), and GNR-gN-HBC in the neutral antiferromagnetic state (c). The isovalue is 0.003~e/Bohr$^3$. Panel (a) is also shown in Figure~3c. The two colors in the spin-density isosurface plots represent spin up and spin down.
}
\label{spin polari} 
\end{center}
\end{figure}
\clearpage

\begin{figure}[!h]
\begin{center}
\includegraphics[width=0.9 \linewidth]{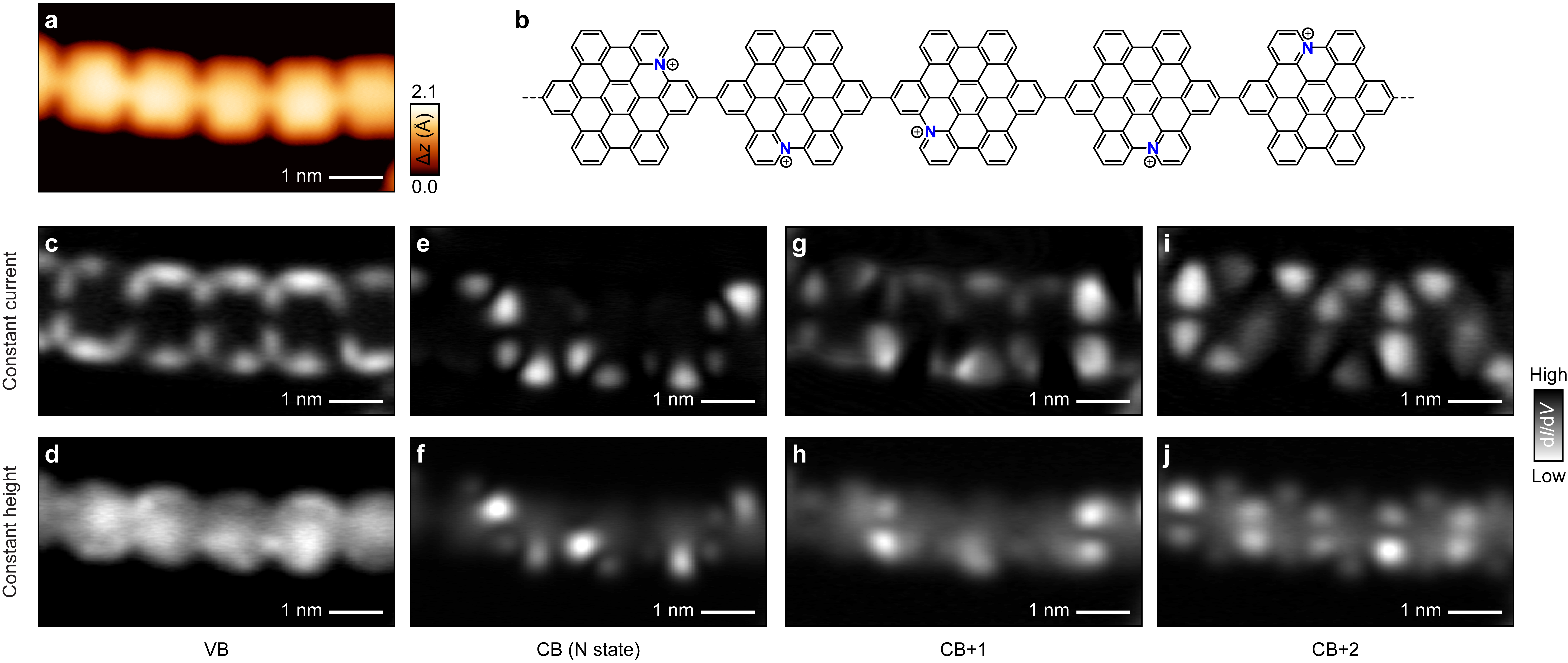}
\caption{Constant-current and constant-height d\textit{I}/d\textit{V} maps of \textit{poly}-gN-HBC. (a) High resolution STM image of a \textit{poly}-gN-HBC segment ($V = -0.10$~V, $I = 200$~pA). (b) Chemical structure of the segment in (a). (c--j) Constant-current (top) and constant-height (bottom) d\textit{I}/d\textit{V} maps of the segment acquired at voltages corresponding to the resonances in Figure~4a: VB (c, d), CB (e, f), CB+1 (g, h) and CB+2 (i, j). Scanning parameters for the constant-current d\textit{I}/d\textit{V} maps: $V = -1.85$~V, $I = 250$~pA (c), $V = 0.70$~V, $I = 200$~pA (e), $V = 1.44$~V, $I = 200$~pA (g), and $V = 1.82$~V, $I = 200$~pA (i). STM set-point for the constant-height d\textit{I}/d\textit{V} maps: $V = -1.85$~V, $I = 300$~pA (d), $V = 0.90$~V, $I = 270$~pA (f), $V = 1.44$~V, $I = 250$~pA (h), and $V = 1.82$~V, $I = 280$~pA (j); $\Delta z = 0$~\AA. The feedback loop was opened with the tip above the center of the third gN-HBC unit at the respective STM set-point currents and voltages, and the scanning voltages are the same as the set-point voltages. For all maps, $V_{\mathrm{rms}} = 16$~mV.}
\label{maps} 
\end{center}
\end{figure}

\bibliography{bibliography}

@article{joucken2019electronic,
  title={Electronic Properties of Chemically Doped Graphene},
  author={Joucken, Fr{\'e}d{\'e}ric and Henrard, Luc and Lagoute, J{\'e}r{\^o}me},
  journal={Physical Review Materials},
  volume={3},
  number={11},
  pages={110301},
  year={2019},
  publisher={APS},
  doi={https://doi.org/10.1103/PhysRevMaterials.3.110301}
}

@article{wang2012review,
  title={Review on Recent Progress in Nitrogen-Doped Graphene: Synthesis, Characterization, and its Potential Applications},
  author={Wang, Haibo and Maiyalagan, Thandavarayan and Wang, Xin},
  journal={ACS Catalysis},
  volume={2},
  number={5},
  pages={781--794},
  year={2012},
  publisher={ACS Publications},
  doi={https://doi.org/10.1021/cs200652y}
}

@article{da2021electronic,
  title={Electronic Properties of N-rich Graphene Nano-Chevrons},
  author={da Costa Azev{\^e}do, Anderson Soares and Saraiva-Souza, Aldilene and Meunier, Vincent and Gir{\~a}o, Eduardo Costa},
  journal={Physical Chemistry Chemical Physics},
  volume={23},
  number={23},
  pages={13204--13215},
  year={2021},
  publisher={Royal Society of Chemistry},
  doi={https://doi.org/10.1039/D1CP00197C}
}

@article{clair2019controlling,
  title={Controlling a Chemical Coupling Reaction on a Surface: Tools and Strategies for On-surface Synthesis},
  author={Clair, Sylvain and de Oteyza, Dimas G},
  journal={Chemical Reviews},
  volume={119},
  number={7},
  pages={4717--4776},
  year={2019},
  publisher={ACS Publications},
  doi={https://doi.org/10.1021/acs.chemrev.8b00601}
}

@article{wang2017exploration,
  title={Exploration of Pyrazine-Embedded Antiaromatic Polycyclic Hydrocarbons Generated by Solution and On-surface Azomethine Ylide Homocoupling},
  author={Wang, Xiao-Ye and Richter, Marcus and He, Yuanqin and Bj{\"o}rk, Jonas and Riss, Alexander and Rajesh, Raju and Garnica, Manuela and Hennersdorf, Felix and Weigand, Jan J and Narita, Akimitsu and Feng, Xinliang and Auwärter, Willi and Barth, Johannes V. and Palma, Carlos-Andres and Müllen, Klaus},
  journal={Nature Communications},
  volume={8},
  number={1},
  pages={1948},
  year={2017},
  publisher={Nature Publishing Group UK London},
  doi={https://doi.org/10.1038/s41467-017-01934-1}
}

@article{piskun2020covalent,
  title={Covalent C--N Bond Formation through a Surface Catalyzed Thermal Cyclodehydrogenation},
  author={Piskun, Ilya and Blackwell, Raymond and Jornet-Somoza, Joaquim and Zhao, Fangzhou and Rubio, Angel and Louie, Steven G and Fischer, Felix R},
  journal={Journal of the American Chemical Society},
  volume={142},
  number={8},
  pages={3696--3700},
  year={2020},
  publisher={ACS Publications},
  doi={https://doi.org/10.1021/jacs.9b13507}
}

@article{sun2021heterocyclic,
  title={Heterocyclic Ring-Opening of Nanographene on Au (111)},
  author={Sun, Kewei and Sugawara, Kazuma and Lyalin, Andrey and Ishigaki, Yusuke and Uosaki, Kohei and Taketsugu, Tetsuya and Suzuki, Takanori and Kawai, Shigeki},
  journal={Angewandte Chemie International Edition},
 volume = {60},
number = {17},
pages = {9427-9432},
  year={2021},
  publisher={Wiley Online Library},
  doi={https://doi.org/10.1002/anie.202017137}
}

@article{wen2022magnetic,
  title={Magnetic Interactions in Substitutional Core-Doped Graphene Nanoribbons},
  author={Wen, Ethan Chi Ho and Jacobse, Peter H and Jiang, Jingwei and Wang, Ziyi and McCurdy, Ryan D and Louie, Steven G and Crommie, Michael F and Fischer, Felix R},
  journal={Journal of the American Chemical Society},
  volume={144},
  number={30},
  pages={13696--13703},
  year={2022},
  publisher={ACS Publications},
  doi={https://doi.org/10.1021/jacs.2c04432}
}

@article{fu2025building,
  title={Building Spin-1/2 Antiferromagnetic Heisenberg Chains with Diaza-Nanographenes},
  author={Fu, Xiaoshuai and Huang, Li and Liu, Kun and Henriques, Jo{\~a}o CG and Gao, Yixuan and Han, Xianghe and Chen, Hui and Wang, Yan and Palma, Carlos-Andres and Cheng, Zhihai and Lin, Xiao and Ma, Ji and Fernández-Rossier, Joaquín and Feng, Xinliang and Gao, Hong-Jun},
  journal={Nature Synthesis},
  pages={1--10},
  year={2025},
   volume={4},
  number={6},
  publisher={Nature Publishing Group UK London},
  doi={https://doi.org/10.1038/s44160-025-00743-5}
}

@article{sun2025surface,
  title={On-surface Synthesis of Heisenberg Spin-1/2 Antiferromagnetic Molecular Chains},
  author={Sun, Kewei and Cao, Nan and Silveira, Orlando J and Fumega, Adolfo O and Hanindita, Fiona and Ito, Shingo and Lado, Jose L and Liljeroth, Peter and Foster, Adam S and Kawai, Shigeki},
  journal={Science Advances},
  volume={11},
  number={9},
  pages={eads1641},
  year={2025},
  publisher={American Association for the Advancement of Science},
  doi={https://doi.org/10.1126/sciadv.ads1641}
}

@article{ketabi2016tuning,
  title={Tuning the Electronic Structure of Graphene through Nitrogen Doping: Experiment and Theory},
  author={Ketabi, Niloofar and de Boer, Tristan and Karakaya, Mehmet and Zhu, Jingyi and Podila, Ramakrishna and Rao, Apparao M and Kurmaev, Ernst Z and Moewes, Alexander},
  journal={RSC Advances},
  volume={6},
  number={61},
  pages={56721--56727},
  year={2016},
  publisher={Royal Society of Chemistry},
  doi={https://doi.org/10.1039/C6RA07546K}
}

@article{shao2010nitrogen,
  title={Nitrogen-Doped Graphene and its Electrochemical Applications},
  author={Shao, Yuyan and Zhang, Sheng and Engelhard, Mark H and Li, Guosheng and Shao, Guocheng and Wang, Yong and Liu, Jun and Aksay, Ilhan A and Lin, Yuehe},
  journal={Journal of Materials Chemistry},
  volume={20},
  number={35},
  pages={7491--7496},
  year={2010},
  publisher={Royal Society of Chemistry},
  doi={https://doi.org/10.1039/C0JM00782J}
}

@article{qu2010nitrogen,
  title={Nitrogen-Doped Graphene as Efficient Metal-Free Electrocatalyst for Oxygen Reduction in Fuel Cells},
  author={Qu, Liangti and Liu, Yong and Baek, Jong-Beom and Dai, Liming},
  journal={ACS Nano},
  volume={4},
  number={3},
  pages={1321--1326},
  year={2010},
  publisher={ACS Publications},
  doi={https://doi.org/10.1021/nn901850u}
}

@article{deng2016review,
  title={Review on Recent Advances in Nitrogen-Doped Carbons: Preparations and Applications in Supercapacitors},
  author={Deng, Yuanfu and Xie, Ye and Zou, Kaixiang and Ji, Xiulei},
  journal={Journal of Materials Chemistry A},
  volume={4},
  number={4},
  pages={1144--1173},
  year={2016},
  publisher={Royal Society of Chemistry},
  doi={https://doi.org/10.1039/C5TA08620E}
}

@article{biswas2021surface,
  title={On-Surface Synthesis of a Dicationic Diazahexabenzocoronene Derivative on the Au (111) Surface},
  author={Biswas, Kalyan and Urgel, Jos{\'e} I and Xu, Kun and Ma, Ji and S{\'a}nchez-Grande, Ana and Mutombo, Pingo and Gallardo, Aurelio and Lauwaet, Koen and Mallada, Benjamin and de la Torre, Bruno and Gallego, José M. and Miranda, Rodolfo and Jelínek, Pavel and Feng, Xinliang and Écija, David},
  journal={Angewandte Chemie International Edition},
  volume={60},
  number={48},
  pages={25551--25556},
  year={2021},
  publisher={Wiley Online Library},
  doi={https://doi.org/10.1002/anie.202111863}
}

@article{wang2022aza,
  title={Aza-Triangulene: On-surface Synthesis and Electronic and Magnetic Properties},
  author={Wang, Tao and Berdonces-Layunta, Alejandro and Friedrich, Niklas and Vilas-Varela, Manuel and Calupitan, Jan Patrick and Pascual, Jose Ignacio and Pe{\~n}a, Diego and Casanova, David and Corso, Martina and de Oteyza, Dimas G},
  journal={Journal of the American Chemical Society},
  volume={144},
  number={10},
  pages={4522--4529},
  year={2022},
  publisher={ACS Publications},
  doi={https://doi.org/10.1021/jacs.1c12618}
}

@article{wen2023fermi,
  title={Fermi-Level Engineering of Nitrogen Core-Doped Armchair Graphene Nanoribbons},
  author={Wen, Ethan Chi Ho and Jacobse, Peter H and Jiang, Jingwei and Wang, Ziyi and Louie, Steven G and Crommie, Michael F and Fischer, Felix R},
  journal={Journal of the American Chemical Society},
  volume={145},
  number={35},
  pages={19338--19346},
  year={2023},
  publisher={ACS Publications},
  doi={https://doi.org/10.1021/jacs.3c05755}
}

@article{vilas2023surface,
  title={On-surface Synthesis and Characterization of a High-Spin Aza-[5]-Triangulene},
  author={Vilas-Varela, Manuel and Romero-Lara, Francisco and Vegliante, Alessio and Calupitan, Jan Patrick and Mart{\'\i}nez, Adri{\'a}n and Meyer, Lorenz and Uriarte-Amiano, Unai and Friedrich, Niklas and Wang, Dongfei and Schulz, Fabian and Koval, Natalia E. and Sandoval-Salinas, María E. and Casanova, David and Corso, Martina and Artacho, Emilio and Peña, Diego and Pascual, José Ignacio},
  journal={Angewandte Chemie International Edition},
  volume={62},
  number={41},
  pages={e202307884},
  year={2023},
  publisher={Wiley Online Library},
  doi={ https://doi.org/10.1002/anie.202307884}
}

@article{willke2015doping,
  title={Doping of Graphene by Low-Energy Ion Beam Implantation: Structural, Electronic, and Transport Properties},
  author={Willke, Philip and Amani, Julian A and Sinterhauf, Anna and Thakur, Sangeeta and Kotzott, Thomas and Druga, Thomas and Weikert, Steffen and Maiti, Kalobaran and Hofs\"ass, Hans and Wenderoth, Martin},
  journal={Nano Letters},
  volume={15},
  number={8},
  pages={5110--5115},
  year={2015},
  publisher={ACS Publications},
  doi={https://doi.org/10.1021/acs.nanolett.5b01280}
}

@article{vegliante2025surface,
  title={On-surface Synthesis of a Ferromagnetic Molecular Spin Trimer},
  author={Vegliante, Alessio and Vilas-Varela, Manuel and Ortiz, Ricardo and Romero Lara, Francisco and Kumar, Manish and G{\'o}mez-Rodrigo, Luc{\'\i}a and Trivini, Stefano and Schulz, Fabian and Soler-Polo, Diego and Ahmoum, Hassan and Frederiksen, Thomas and Jelínek, Pavel and Pascual, Jose Ignacio and Peña, Diego},
  journal={Journal of the American Chemical Society},
  year={2025},
  volume={147},
  pages={19530--19538},
  number={23},
  publisher={ACS Publications},
  doi={https://doi.org/10.1021/jacs.4c15736}
}

@article{li2024surface,
  title={On-surface Synthesis of Triaza[5]triangulene through Cyclodehydrogenation and its Magnetism},
  author={Li, Donglin and Silveira, Orlando J and Matsuda, Takuma and Hayashi, Hironobu and Maeda, Hiromitsu and Foster, Adam S and Kawai, Shigeki},
  journal={Angewandte Chemie International Edition},
  volume={63},
  number={45},
  pages={e202411893},
  year={2024},
  publisher={Wiley Online Library},
  doi={https://doi.org/10.1002/anie.202411893}
}

@article{gao2022selective,
  title={Selective Activation of Four Quasi-Equivalent C--H Bonds Yields N-Doped Graphene Nanoribbons with Partial Corannulene Motifs},
  author={Gao, Yixuan and Huang, Li and Cao, Yun and Richter, Marcus and Qi, Jing and Zheng, Qi and Yang, Huan and Ma, Ji and Chang, Xiao and Fu, Xiaoshuai and Palma, Carlos-Andres and Cheng, Zhihai and Feng, Xinliang and Du, Shixuan and Gao, Hong-Jun},
  journal={Nature Communications},
  volume={13},
  number={1},
  pages={6146},
  year={2022},
  publisher={Nature Publishing Group UK London},
  doi={https://doi.org/10.1038/s41467-022-33898-2}
}

@article{guo2010controllable,
  title={Controllable N-Doping of Graphene},
  author={Guo, Beidou and Liu, Qian and Chen, Erdan and Zhu, Hewei and Fang, Liang and Gong, Jian Ru},
  journal={Nano Letters},
  volume={10},
  number={12},
  pages={4975--4980},
  year={2010},
  publisher={ACS Publications},
  DOI={https://doi.org/10.1021/nl103079j},
}

@article{joucken2012localized,
  title={Localized State and Charge Transfer in Nitrogen-Doped Graphene},
  author={Joucken, Fr{\'e}d{\'e}ric and Tison, Yann and Lagoute, J{\'e}r{\^o}me and Dumont, Jacques and Cabosart, Damien and Zheng, Bing and Repain, Vincent and Chacon, Cyril and Girard, Yann and Botello-M{\'e}ndez, Andr{\'e}s Rafael and Rousset, Sylvie and Sporken, Robert and Charlier, Jean-Christophe and Henrard, Luc},
  journal={Physical Review B},
  volume={85},
  number={16},
  pages={161408},
  year={2012},
  publisher={APS},
  doi={https://doi.org/10.1103/PhysRevB.85.161408}
}

@article{joucken2015charge,
  title={Charge Transfer and Electronic Doping in Nitrogen-Doped Graphene},
  author={Joucken, Fr{\'e}d{\'e}ric and Tison, Yann and Le F{\`e}vre, Patrick and Tejeda, Antonio and Rousset, Sylvie and  Ghijsen, Jacques and Sporken, Robert and Amara, Hakim and Ducastelle, François and Lagoute, Jérôme},
  journal={Scientific Reports},
  volume={5},
  number={1},
  pages={14564},
  year={2015},
  publisher={Nature Publishing Group UK London},
  doi={https://doi.org/10.1038/srep14564}
}

@article{bassi2024preferential,
  title={Preferential Graphitic-Nitrogen Formation in Pyridine-Extended Graphene Nanoribbons},
  author={Bassi, Nicol{\`o} and Xu, Xiushang and Xiang, Feifei and Krane, Nils and Pignedoli, Carlo A and Narita, Akimitsu and Fasel, Roman and Ruffieux, Pascal},
  journal={Communications Chemistry},
  volume={7},
  number={1},
  pages={274},
  year={2024},
  doi={https://doi.org/10.1038/s42004-024-01344-7}
}

@article{zhang2024synthesis,
  title={Synthesis of Azahexabenzocoronenium Salts through a Formal [3+3] Cycloaddition Strategy},
  author={Zhang, Xinjiang and Li, Donglin and Tan, Cheryl Cai Hui and Hanindita, Fiona and Hamamoto, Yosuke and Foster, Adam S and Kawai, Shigeki and Ito, Shingo},
  journal={Nature Synthesis},
  volume={3},
  number={10},
  pages={1283--1291},
  year={2024},
  publisher={Nature Publishing Group UK London},
  doi={https://doi.org/10.1038/s44160-024-00595-5}
}

@article{cai_atomically_2010,
    title = {Atomically Precise Bottom-up Fabrication of Graphene Nanoribbons},
    volume = {466},
    issn = {1476-4687},
    url = {https://doi.org/10.1038/nature09211},
    doi = {https://doi.org/10.1038/nature09211},
    language = {en},
    number = {7305},
    urldate = {2022-02-25},
    journal = {Nature},
    author = {Cai, Jinming and Ruffieux, Pascal and Jaafar, Rached and Bieri, Marco and Braun, Thomas and Blankenburg, Stephan and Muoth, Matthias and Seitsonen, Ari P. and Saleh, Moussa and Feng, Xinliang and Müllen, Klaus and Fasel, Roman},
    month = jul,
    year = {2010},
    pages = {470--473},
}

@article{friedman2016electronic,
  title={Electronic Transport and Localization in Nitrogen-Doped Graphene Devices using Hyperthermal Ion Implantation},
  author={Friedman, Adam L and Cress, Cory D and Schmucker, Scott W and Robinson, Jeremy T and van ‘t Erve, Olaf MJ},
  journal={Physical Review B},
  volume={93},
  number={16},
  pages={161409},
  year={2016},
  publisher={APS},
  doi={https://doi.org/10.1103/PhysRevB.93.161409}
}

@article{houtsma2021atomically,
  title={Atomically Precise Graphene Nanoribbons: Interplay of Structural and Electronic Properties},
  author={Houtsma, RS Koen and de la Rie, Joris and St{\"o}hr, Meike},
  journal={Chemical Society Reviews},
  volume={50},
  number={11},
  pages={6541--6568},
  year={2021},
  publisher={Royal Society of Chemistry},
  doi={https://doi.org/10.1039/D0CS01541E}
}

@article{calupitan2023emergence,
  title={Emergence of $\pi$-Magnetism in Fused Aza-Triangulenes: Symmetry and Charge Transfer Effects},
  author={Calupitan, Jan Patrick and Berdonces-Layunta, Alejandro and Aguilar-Galindo, Fernando and Vilas-Varela, Manuel and Pe{\~n}a, Diego and Casanova, David and Corso, Martina and de Oteyza, Dimas G and Wang, Tao},
  journal={Nano Letters},
  volume={23},
  number={21},
  pages={9832--9840},
  year={2023},
  publisher={ACS Publications},
  doi={https://doi.org/10.1021/acs.nanolett.3c02586}
}

@article{ooeTwoStepOnSurfaceSynthesis2023,
  title = {Two-{{Step On-Surface Synthesis}} of {{One-Dimensional Nanographene Chains}}},
  author = {Ooe, Hiroaki and Ikeda, Kaede and Yokoyama, Takashi},
  year = {2023},
  month = apr,
  journal = {The Journal of Physical Chemistry C},
  volume = {127},
  number = {16},
  pages = {7659--7665},
  publisher = {American Chemical Society},
  issn = {1932-7447},
  doi={https://doi.org/10.1021/acs.jpcc.3c00373}
}

@article{kawai2018multiple,
  title={Multiple Heteroatom Substitution to Graphene Nanoribbon},
  author={Kawai, Shigeki and Nakatsuka, Soichiro and Hatakeyama, Takuji and Pawlak, R{\'e}my and Meier, Tobias and Tracey, John and Meyer, Ernst and Foster, Adam S},
  journal={Science Advances},
  volume={4},
  number={4},
  pages={eaar7181},
  year={2018},
  publisher={American Association for the Advancement of Science},
  doi={https://doi.org/10.1126/sciadv.aar7181}
}

@article{zhao2012production,
  title={Production of Nitrogen-Doped Graphene by Low-Energy Nitrogen Implantation},
  author={Zhao, Wei and H\"ofert, O and Gotterbarm, Karin and Zhu, JF and Papp, C and Steinr\"uck, H-P},
  journal={The Journal of Physical Chemistry C},
  volume={116},
  number={8},
  pages={5062--5066},
  year={2012},
  publisher={ACS Publications},
  doi={https://doi.org/10.1021/jp209927m}
}

@article{cai2014graphene,
  title={Graphene Nanoribbon Heterojunctions},
  author={Cai, Jinming and Pignedoli, Carlo A and Talirz, Leopold and Ruffieux, Pascal and S{\"o}de, Hajo and Liang, Liangbo and Meunier, Vincent and Berger, Reinhard and Li, Rongjin and Feng, Xinliang and M{\"u}llen, Klaus and Fasel, Roman},
  journal={Nature Nanotechnology},
  volume={9},
  number={11},
  pages={896--900},
  year={2014},
  publisher={Nature Publishing Group},
  doi={https://doi.org/10.1038/nnano.2014.184}
}

@article{bronner_aligning_2013,
    title = {Aligning the {Band} {Gap} of {Graphene} {Nanoribbons} by {Monomer} {Doping}},
    volume = {52},
    issn = {1521-3773},
    url = {https://onlinelibrary.wiley.com/doi/abs/10.1002/anie.201209735},
  doi={https://doi.org/10.1002/anie.201209735},
    number = {16},
    urldate = {2022-02-25},
    journal = {Angewandte Chemie International Edition},
    author = {Bronner, Christopher and Stremlau, Stephan and Gille, Marie and Brauße, Felix and Haase, Anton and Hecht, Stefan and Tegeder, Petra},
    year = {2013},
    pages = {4422--4425},
}

@article{yakutovich2021aiidalab,
  title={AiiDAlab--an Ecosystem for Developing, Executing, and Sharing Scientific Workflows},
  author={Yakutovich, Aliaksandr V and Eimre, Kristjan and Sch{\"u}tt, Ole and Talirz, Leopold and Adorf, Carl S and Andersen, Casper W and Ditler, Edward and Du, Dou and Passerone, Daniele and Smit, Berend and Marzari, Nicola and Pizzi, Giovanni and Pignedoli, Carlo A.},
  journal={Computational Materials Science},
  volume={188},
  pages={110165},
  year={2021},
  publisher={Elsevier},
  doi={https://doi.org/10.1016/j.commatsci.2020.110165}
}

@article{pizzi2016aiida,
  title={AiiDA: Automated Interactive Infrastructure and Database for Computational Science},
  author={Pizzi, Giovanni and Cepellotti, Andrea and Sabatini, Riccardo and Marzari, Nicola and Kozinsky, Boris},
  journal={Computational Materials Science},
  volume={111},
  pages={218--230},
  year={2016},
  publisher={Elsevier},
  doi={https://doi.org/10.1016/j.commatsci.2015.09.013}
}

@article{hutter2014cp2k,
  title={CP2K: Atomistic Simulations of Condensed Matter Systems},
  author={Hutter, J{\"u}rg and Iannuzzi, Marcella and Schiffmann, Florian and VandeVondele, Joost},
  journal={WIREs Computational Molecular Science},
  volume={4},
  number={1},
  pages={15--25},
  year={2014},
  publisher={Wiley Online Library},
  doi={https://doi.org/10.1002/wcms.1159}
}

@article{giannozzi2009quantum,
  title={QUANTUM ESPRESSO: a Modular and Open-Source Software Project for Quantum Simulations of Materials},
  author={Giannozzi, Paolo and Baroni, Stefano and Bonini, Nicola and Calandra, Matteo and Car, Roberto and Cavazzoni, Carlo and Ceresoli, Davide and Chiarotti, Guido L and Cococcioni, Matteo and Dabo, Ismaila and Dal Corso, Andrea and de Gironcoli, Stefano and Fabris, Stefano and Fratesi, Guido and Gebauer, Ralph and Gerstmann, Uwe and Gougoussis, Christos and Kokalj, Anton and Lazzeri, Michele and Martin-Samos, Layla and Marzari, Nicola and Mauri, Francesco and Mazzarello, Riccardo and Paolini, Stefano and Pasquarello, Alfredo and Paulatto, Lorenzo and Sbraccia, Carlo and Scandolo, Sandro and Sclauzero, Gabriele and Seitsonen, Ari P and Smogunov, Alexander and Umari, Paolo and Wentzcovitch, Renata M},
  journal={Journal of Physics: Condensed Matter},
  volume={21},
  number={39},
  pages={395502},
  year={2009},
  publisher={IOP Publishing},
  doi={https://doi.org/10.1088/0953-8984/21/39/395502}
}

@article{perdew1996generalized,
  title={Generalized Gradient Approximation Made Simple},
  author={Perdew, John P and Burke, Kieron and Ernzerhof, Matthias},
  journal={Physical Review Letters},
  volume={77},
  number={18},
  pages={3865},
  year={1996},
  publisher={APS},
  doi={https://doi.org/10.1103/PhysRevLett.77.3865}
}

@article{grimme2010consistent,
  title={A Consistent and Accurate ab initio Parametrization of Density Functional Dispersion Correction (DFT-D) for the 94 Elements H-Pu},
  author={Grimme, Stefan and Antony, Jens and Ehrlich, Stephan and Krieg, Helge},
  journal={The Journal of Chemical Physics},
  volume={132},
  number={15},
  pages={154104},
  year={2010},
  doi={https://doi.org/10.1063/1.3382344},
  publisher={AIP Publishing}
}

@article{hapala2014mechanism,
  title={Mechanism of High-Resolution STM/AFM Imaging with Functionalized Tips},
  author={Hapala, Prokop and Kichin, Georgy and Wagner, Christian and Tautz, F Stefan and Temirov, Ruslan and Jel{\'\i}nek, Pavel},
  journal={Physical Review B},
  volume={90},
  number={8},
  pages={085421},
  year={2014},
  publisher={APS},
  doi={https://doi.org/10.1103/PhysRevB.90.085421}
}

@article{Giessibl,
    author = {Giessibl, Franz J.},
    title = {High-Speed Force Sensor for Force Microscopy and Profilometry Utilizing a Quartz Tuning Fork},
    journal = {Applied Physics Letters},
    volume = {73},
    number = {26},
    pages = {3956-3958},
    year = {1998},
    month = {12},
    issn = {0003-6951},
    doi = {https://doi.org/10.1063/1.122948},
    url = {https://doi.org/10.1063/1.122948},
}

@article{Albrecht,
    author = {Albrecht, T. R. and Grütter, P. and Horne, D. and Rugar, D.},
    title = {Frequency Modulation Detection using high‐Q Cantilevers for Enhanced Force Microscope Sensitivity},
    journal = {Journal of Applied Physics},
    volume = {69},
    number = {2},
    pages = {668-673},
    year = {1991},
    month = {01},
    issn = {0021-8979},
    doi = {https://doi.org/10.1063/1.347347},
}

@article{ZhiHao,
author = {Li, Zhi-Hao and Dai, Jia-Qi and Luo, Guan and Li, Ruo-Ning and Zhao, An-Jing and Duan, Jun-Jie and Ge, Yu and Wang, Zi-Cong and Ji, Wei and Chen, Ting and Wang, Dong and Wan, Li-Jun},
title = {Atomic-Precision Engineering and Visualizing of Chiral Electronic States in Nitrogen-Doped Nanographenes},
journal = {Angewandte Chemie International Edition},
volume = {65},
number = {24},
pages = {e2278978},
doi = {https://doi.org/10.1002/anie.2278978},
url = {https://onlinelibrary.wiley.com/doi/abs/10.1002/anie.2278978},
eprint = {https://onlinelibrary.wiley.com/doi/pdf/10.1002/anie.2278978},
year = {2026}
}

@article{saleh2010triphenylene,
  title={Triphenylene-Based Polymers for Blue Polymeric Light Emitting Diodes},
  author={Saleh, Moussa and Baumgarten, Martin and Mavrinskiy, Alexey and Sch\"afer, Thomas and M\"ullen, Klaus},
  journal={Macromolecules},
  volume={43},
  number={1},
  pages={137--143},
  year={2010},
  publisher={ACS Publications},
  doi = {https://doi.org/10.1021/ma901912t},
}

@article{gholinejad2016palladium,
  title={Palladium Supported on Bis(indolyl)methane Functionalized Magnetite Nanoparticles as an Efficient Catalyst for Copper-Free Sonogashira-Hagihara Reaction},
  author={Gholinejad, Mohammad and Neshat, Abdollah and Zareh, Fatemeh and Najera, Carmen and Razeghi, Mehran and Khoshnood, Abbas},
  journal={Applied Catalysis A: General},
  volume={525},
  pages={31--40},
  year={2016},
  publisher={Elsevier},
  doi = {https://doi.org/10.1016/j.apcata.2016.06.041}
}

@Article{yakutovich2026,
author ="Yakutovich, Aliaksandr V. and Hollas, Daniel and Bainglass, Edan and Yu, Jusong and Battaglia, Corsin and Bonacci, Miki and Fernandez Vilanova, Lucas and Henne, Stephan and Kaestner, Anders and Kenzelmann, Michel and Kimbell, Graham and Lass, Jakob and Lopes, Fabio and Mazzone, Daniel G. and Ortega-Guerrero, Andres and Wang, Xing and Marzari, Nicola and Pignedoli, Carlo A. and Pizzi, Giovanni",
title  ="Accelerating discovery across scientific disciplines through reproducible workflows with AiiDAlab",
journal  ="Digital Discovery",
year  ="2026",
volume  ="5",
issue  ="5",
pages  ="2310-2324",
publisher  ="RSC",
doi  ="10.1039/D5DD00567A",
url  ="http://dx.doi.org/10.1039/D5DD00567A",
}

\end{document}